\documentclass[nofootinbib,aps,superscriptaddress,11pt]{revtex4-1}
\usepackage{geometry}
\usepackage{graphicx}
\usepackage{dcolumn}
\usepackage{bm}
\usepackage{epsfig,amsmath}
\usepackage{amssymb}
\usepackage{subfigure,esint}
\usepackage{float}
\usepackage[usenames,dvipsnames]{color}
\usepackage[pagebackref=false, colorlinks=true]{hyperref}
\definecolor{redish}{rgb}{0.7,0.2,0.0}  
\definecolor{bluish}{rgb}{0.2,0.5,0.8}

\hypersetup{linkcolor=redish,          
	citecolor=blue,        
	filecolor=magenta,      
	urlcolor=blue}          

\begin{document}
\author{Kunal Pal}\email{kunalpal@iitk.ac.in}
\affiliation{
	Department of Physics, Indian Institute of Technology Kanpur, \\ Kanpur 208016, India}
\author{Kuntal Pal}\email{kuntal@iitk.ac.in}
\affiliation{
	Department of Physics, Indian Institute of Technology Kanpur, \\ Kanpur 208016, India}
\author{Pratim Roy}\email{pratimroy@hri.res.in}
\affiliation{Harish-Chandra Research Institute, Chhatnag Road, Jhunsi,
	Prayagraj 211019, India}
\affiliation{
	Homi Bhabha National Institute, Training School Complex,
	Anushaktinagar, Mumbai 400094, India}
\author{Tapobrata Sarkar}\email{tapo@iitk.ac.in}
\affiliation{
	Department of Physics, Indian Institute of Technology Kanpur, \\ Kanpur 208016, India}
\title{\Large Regularising the JNW and JMN naked singularities}

\begin{abstract}
We extend the method of Simpson and Visser (SV) of regularising a black hole spacetime,
to cases where the initial metric represents a globally naked singularity. We choose two particular geometries,
the Janis-Newman-Winicour (JNW) metric representing the solution of an Einstein-scalar field system, and the  
Joshi-Malafarina-Narayan (JMN) metric that represents the asymptotic equilibrium configuration of a collapsing 
star supported by  tangential pressures as the starting configuration. We illustrate several novel features for
the modified versions of the JNW and JMN spacetimes. In particular, we show that, depending on the
values of the parameters involved the modified JNW metric may represents either a two way traversable wormhole
or it may retain the original naked singularity.
On the other hand, the SV modified JMN geometry is always a wormhole. 
Particle motion and observational aspects of these new geometries are investigated and are shown to posses
interesting features. We also study the quasinormal modes of different branches of the regularised spacetime and explore their stability properties.
\end{abstract}

\maketitle	
\section{Introduction}
\label{intro}
	
Despite being the most successful theory of gravity till date, it is widely believed that general relativity (GR) \cite{carroll}
may break down in high curvature regimes, where it will be replaced by an yet unknown theory of quantum gravity. 
For this reason, the final fate of a gravitational collapse, starting from some well behaved set of initial data 
is still an open problem, as the analysis carried out in the classical (or effectively semiclassical) theory 
may received important quantum corrections in the strong gravity regime. In this context, two important questions are known
to arise. First, if the regular interior of a collapsing star in realistic situations produce a spacetime singularity, 
which is specified by a diverging curvature scalar, and secondly, whether this singularity is visible to an asymptotic observer. 
The latter is thought to be prohibited by the cosmic censorship conjecture (CCC) of Penrose \cite{RP1}, which is 
yet to be proved in full generality. In the absence of a fully accepted theory of quantum gravity, a popular line of 
research to study the above mentioned problems is to construct a model of `regular black holes,' where the singularity 
is replaced by a region of regular curvature. On the other hand, the notion of a `horizonless compact object' 
with or without a singularity, like a wormhole (WH) \cite{visser,damour}, 
a naked singularity (NS) etc., has also gained considerable attention 
in the literature. Although the presence of a NS without an event horizon would violate the CCC, there are 
well studied models in the literature, where the end states of a viable collapse process produces a NS, 
and therefore it is fair to say that a firm conclusion is yet to be reached in the matter.

Starting from the seminal work of Bardeen \cite{JMB}, there are abundant models of regular black holes 
which are constructed either from a purely phenomenological point of view, or as solutions of an effective 
semiclassical gravity theory (for a review and further references, see \cite{HM}). The basic idea of these kinds of 
phenomenological models is to replace the Schwarzschild mass by some general family of mass functions, that would 
render the curvature scalars finite everywhere in the spacetime. Though this looks promising at first sight, 
there are many models of regular black holes that have an inner Cauchy horizon, and its instability under 
mass inflation can be problematic for realistic model building (see \cite{Casadio1} and references therein). 
In view of the above discussions, we are led to consider an interesting one parameter class of models 
for regularising the Schwarzschild singularity, that was 
proposed recently by Simpson and Visser (SV) \cite{SV1}. Here, the central singularity was replaced by a regular region, 
that can be timelike, null, or spacelike, depending on the parameter involved in the geometry, and the 
global geometry represents either a traversable WH or a regular BH with one horizon. Such models, dubbed as `black-bounce' 
spacetimes, not only resolve the Schwarzschild singularity, but also provide a link between two different 
classes of spacetime structures, namely the BH and the WH, and therefore is a promising way of further constructing 
regular black holes and resolving the singularity problem of classical GR. See \cite{franzin2} - \cite{Barrientos:2022avi} 
for a selection of recent works in this direction.	

  In view of the above discussions, an interesting question concerns the behaviour of globally naked singularities that 
can be regularised following the SV procedure. It is this issue that we address in this paper.  
Here, our purpose is to extend the method of SV and construct regular solutions starting from a globally 
naked singularity, and in the process unify two different classes of metrics, 
namely the WH and NS. To this end, we first apply the SV regularisation 
scheme to the Janis-Newman-Winicour (JNW) \cite{JNW, Virbhadra1} class of spacetimes which have a naked curvature singularity 
at a finite radial location, and reduces to the Schwarzschild black hole for a particular limit of the parameters involved. 
We construct a deformed JNW metric that interpolates between a BH, WH and NS for different ranges of the parameters.  
As a second example, we construct a similar deformation of the Joshi-Malafarina-Narayan (JMN) \cite{JMN} spacetime, that 
contains a central NS and is known to be a result of gravitational collapse of regular initial data. 
In the following sections, we discuss the resulting metrics and show that these are indeed rich in terms of  global spacetime structures 
and observational aspects. Further, we discuss the quasinormal modes of the regularised JNW metric using the WKB approach 
\cite{Iyer:1986np, Iyer:1986nq} concentrating on the wormhole branch, and comment on the stability of the spacetime, particularly when 
approaching the NS branch by changing the SV parameter.

We always work in units where the Newton's gravitational constant and the speed of light are set to unity.

\noindent
{\bf Note Added :} While this draft was being finalised, the recent paper \cite{Bronnikov} appeared, which
has partial overlap with the results presented in section \ref{JNWsection} of this work. However, the final metric considered in \cite{Bronnikov} is slightly different than that we considered due to a different coordinate system chosen.

\section{Deformed Janis-Newman-Winicour solution}
\label{JNWsection}

The  Janis-Newman-Winicour metric \cite{JNW, Virbhadra1} is a static, spherically symmetric solution 
of the Einstein field equations in the presence of a minimally coupled scalar field, and is given by the line element 
\begin{equation}\label{JNW}
	ds^2=-\Big(1-\frac{b}{r}\Big)^\gamma \text{d}t^2 + \Big(1-\frac{b}{r}\Big)^{-\gamma} \text{d}r^2 + 
	\Big(1-\frac{b}{r}\Big)^{1-\gamma}r^2 \text{d}\Omega^2~.
\end{equation}
The parameters $\gamma$ and $b$  appearing in the JNW metric are related to the ADM mass $M$ and the scalar charge $q$  
by the relations
\begin{equation}\label{parameters}
	\gamma = \frac{2M}{b}~,~~b=2\sqrt{M^2+q^2}~.
\end{equation}
In the limit of vanishing scalar charge $q\rightarrow 0$, the JNW solution reduces to the Schwarzschild solution. 
In the general case, with $q\neq 0$, and real, it can be seen from Eq. (\ref{parameters}) that $\gamma<1$. 
This fact also indicates that the JNW metric is only valid up to the coordinate location $r=b$. It can also be checked 
that for $\gamma < 1$, this metric has a curvature singularity at $r=b$, and since the metric does not possess any event 
horizon, it is a naked singularity \cite{Virbhadra2}. On the other hand, when the scalar charge $q$ is imaginary
such that $\gamma > 1$, this metric represents 
a symmetric traversable WH with two asymptotically flat regions \cite{nandi1, nandi2}.	
	
	Now we apply the SV method via the replacement $r\to\sqrt{r^2+c^2}$, without modifying the form $\text{d}r$ (in which case it will
simply be an uninteresting overall coordinate transformation) in the JNW solution, and the resulting metric becomes 
\begin{equation}\label{RJNW}
	ds^2=-\Big(1-\frac{b}{\sqrt{r^2+c^2}}\Big)^\gamma \text{d}t^2 + \Big(1-\frac{b}{\sqrt{r^2+c^2}}\Big)^{-\gamma} \text{d}r^2 
	+ \Big(1-\frac{b}{\sqrt{r^2+c^2}}\Big)^{1-\gamma}\Big(r^2+c^2\Big) \text{d}\Omega^2~.
\end{equation}
Here $c$ (not to be confused with the speed of light), the SV parameter, is a real and positive quantity having dimensions of length.
The range of coordinates, other than that of $r$ are same as before, and the range of $r$ depends upon the relative 
values of $c$ and $b$, as we explain below in detail. Also, it can readily be seen that the original 
symmetries and the asymptotic properties of the JNW  metric are preserved with the new one. We will get back the 
JNW metric in the limit $c \rightarrow 0$ and the SV metric (i.e., the deformed Schwarzschild metric) 
in the limit $\gamma \rightarrow 1$. 	

Now we elaborate upon the metric of Eq. (\ref{RJNW}) (which we shall call the SV-JNW solution from now on) for 
different parameter ranges. For this, it will be useful to define a coordinate transformation $k^2=r^2+c^2$, doing which we get 
\begin{equation}\label{RJNW2}
	ds^2=-\Big(1-\frac{b}{k}\Big)^\gamma \text{d}t^2 + \Big(1-\frac{b}{k}\Big)^{-\gamma}\frac{k^2}{k^2-c^2} \text{d}k^2 
	+ \Big(1-\frac{b}{k}\Big)^{1-\gamma}k^2 \text{d}\Omega^2~.
\end{equation}
To gain more insight into the structure of the metric, we use another coordinate transformation
\begin{equation}\label{ct2}
	R^2=\Big(1-\frac{b}{k}\Big)^{1-\gamma}k^2~,
\end{equation}
after which the final form of the metric is, 
\begin{equation}\label{metrictR}
	ds^2=-\Big(1-\frac{b}{k(R)}\Big)^\gamma \text{d}t^2 + \Big(1-\frac{b}{k(R)}\Big)^{-\gamma}\frac{k(R)^2}{k(R)^2-c^2} 
	\left(\frac{1}{D(k)^2\Big(1+\frac{b(1-\gamma)}{2(k-b)}\Big)^2}\right)\text{d}R^2 + R^2 \text{d}\Omega^2~,
\end{equation}
and we have defined $D(k)=\Big(1-\frac{b}{k}\Big)^{\frac{1-\gamma}{2}}$. Since Eq.  (\ref{ct2}) can be inverted in
principle to obtain $k$ in terms of the new variable $R$, we have retained its implicit dependence.
To gain insights into this metric, we now compare this to the canonical Morris-Thorne WH \cite{morris}
\begin{equation}
	ds^2=-\exp(2\Phi) \text{d}t^2 + \frac{1}{1-\frac{B(R)}{R}}\text{d}R^2 + R^2 \text{d}\Omega^2~,
\end{equation} 
where $B(R)$ is the shape function and $\Phi(R)$ is the redshift function.
For our metric the shape function is given by
\begin{equation}
	B(k(R))=R\Bigg[1-\Big(1-\frac{b}{k}\Big)\Big(1-\frac{c^2}{k^2}\Big)\Bigg(1+\frac{b(1-\gamma)}{2(k-b)}\Bigg)\Bigg]~.
\end{equation}
 From the form of the redshift function $g_{tt}$ in Eq. (\ref{RJNW2}),  we can see that it vanishes at $k_{1}=b$. 
 Similarly, The $g^{RR}$ component of the above metric in Eq. (\ref{metrictR}) has two zeros, respectively at $k_{2}=c$, 
 and $~k_{3}=\frac{b}{2}(\gamma+1)$. Furthermore, the equation for the radial  null curves for the metrics in 
 the $R, t$ coordinates are given by
\begin{equation}
\frac{dR}{dt}=\Big[\Big(1-\frac{b}{k}\Big)^{\gamma}\frac{(k^2-c^2)\Big(2(k-b)+b(1-\gamma)\Big)}{4k^3(k-b)}\Big]^{1/2}~~.
\end{equation}

Thus, depending upon the nature of the parameters involved in the metric, we can have several interesting cases here, which
we now detail.  First we shall consider the cases with $\gamma<1$.	
	 
	 \noindent
	$\bullet~\gamma <1$ and $b<c$ :
	Here, $k_{3}<k_1$ and hence the root $k_3$ does not play any significant role, 
	since the metric is not valid upto the coordinate value corresponding to $k_3$.  
	In this case, it can readily be seen that the original domain of validity of the JNW metric (up to  $r \geq b$) 
	has changed, and the metric can be continued from $r = \infty$ to $r =-\infty$, through $r=0$. The 
	metric then represents a traversable wormhole, with a timelike throat at $r=0$, connecting two asymptotically flat regions. 
	This is very different from the usual JNW solution, where no WH branch exists for $\gamma < 1$. 
	For completeness, we explicitly check that the conditions for a traversable wormhole \cite{visser} are satisfied here :\\
	(i) Absence of horizons - This can be seen from the metric written in Eq. (\ref{RJNW2}), from which it is evident that the 
	spacetime does not contain any horizon as long as $b <c$ is satisfied.\\
	(ii) Flaring out condition - This condition satisfied by  a traversable wormhole embodies the fact that the 
	throat is the location of minimum area (see \cite{simpsonthesis} for a detailed discussion). 
	Mathematically, the flaring out condition implies that 
	$A^{\prime}(r=r_{0})=0, A^{\prime\prime}(r=r_{0})>0$, where the prime denotes a derivative with respect to the radial coordinate.  
	For the metric of Eq. (\ref{RJNW}), we note that the area of the two sphere is 
	$A(r)=\Big(1-\frac{b}{\sqrt{r^2+c^2}}\Big)^{1-\gamma}(r^2+c^2)$, and hence $A^{\prime}(r=r_{0})=0 \implies r_0=0$, i.e.,  
	we have a stationary point of the area of the spherical surface at the throat $r=r_{0}=0$.  
	To check that this represents a minimum we only need to note that 
	$A^{\prime \prime}(r=r_{0})= \frac{2\big(2c-b(1+\gamma)\big)}{c(1-\frac{b}{c})^{\gamma}}$.
	Thus  as long as the condition $b<c$ satisfied  $A^{\prime \prime}(r=r_{0})>0$, and we see that the area is 
	indeed minimum at the location of the throat, and that the flaring out condition is indeed satisfied.\\
	(iii) Non-singular nature - To show that the resulting geometry is non singular everywhere - which is required for the metric 
	to represent a wormhole geometry, we calculate the curvature scalars. The Ricci scalar in this case  
	$R \sim (1-\frac{b}{\sqrt{r^2+c^2}})^{-2}$, from where we see that it is indeed finite everywhere,
	including $r=0$ and $r=b$. Similarly, the Kretschmann scalar  
	$K \sim (1-\frac{b}{\sqrt{r^2+c^2}})^{-4}$ which is also finite everywhere. 
	
	\noindent
	$\bullet~\gamma <1$ and $b>c$ :
	Here, the range of the radial coordinate is restricted to $\sqrt{b^2-c^2}<r<\infty$, a condition analogous to that in the original 
	JNW metric. The Ricci scalar, given by $R \sim (1-\frac{b}{\sqrt{r^2+c^2}})^{-2}$,  diverges 
	in the limit $r\rightarrow \sqrt{b^2-c^2}$ (the same conclusion can be shown to be valid for the Krestschmann scalar curvature). 
	On the other hand, from the expression of the metric, it can be seen that there is no possibility an event horizon  at the location 
	of the curvature singularity. So, as in the case of the original JNW metric \cite{Virbhadra2}, we conclude that 
	the singularity is not covered by that of an event horizon from a distant observer, and that the spacetime 
	represents a naked singularity for the above parameter ranges. 
	
	\noindent
	$\bullet~\gamma <1$ and $b=c$ :
	This is the limiting case of the two earlier ones. As explained above, here the metric consists of a 
	throat (minimum of the local area), and the curvature singularity, all at the location 
	$r=0$, as well as there is no possibility of an event horizon. So the metric represents a one way wormhole with a singular throat. 
	
	Now we examine the solutions for the range $\gamma>1$.
	This case is similar to that of the original JNW WH \cite{nandi1,nandi2}, although due to the introduction 
	of the the parameter $c$, we will have a richer geometric structure. First note that here $k_{3}>k_{1}$, and in the original 
	coordinate $r$, the three locations $k_i$s are written as 
	\begin{equation}\label{roots}
		r_{1}=\sqrt{b^2-c^2}~,~~ r_{2}=0~,~~ r_{3}=\sqrt{\frac{b^2}{4}(\gamma+1)^2-c^2}~.
	\end{equation}  
	As before we will consider two cases $b>c$ and $b<c$ and their limiting case. 
	
	\noindent
	$\bullet~b<\frac{b}{2}(\gamma+1)<c$ :
	Here, we consider only $r_2=0$ from Eq. (\ref{roots}), which is the only real one in this case.
	 Again, we see that the region of validity of the 
	metric is changed to the range from $r=-\infty$ to $r = \infty$. In this case, the metric represents a traversable wormhole 
	with a throat at $r=0$. One can check the validity of the WH features here, but we omit the details for brevity. 
	
	\noindent
	$\bullet~b<c<\frac{b}{2}(\gamma+1)$ :
	In this case, the value of $\gamma$ is such that only $r_3$ in Eq. (\ref{roots}) is real. 
	We can immediately see that this metric represents a traversable WH with a throat at $r_{3}$, with similar arguments as before. 
	
	Finally note that there is no possibility of any NS branch here, as even in the case $c<b<\frac{b}{2}(\gamma+1)$, 
	the metric represents a WH, since the metric is still valid upto $r_{3}$. So we can conclude that the 
	the NS is fully regularised here, following the SV method. 
	
	\subsection{Photon motion in SV-JNW background}

Now we discuss the motion of a photon in the SV-JNW background, and the resulting shadow structure of the metric 
in Eq. (\ref{RJNW}). Since the metric is spherically symmetric, the structure of the null geodesics is fairly simple, 
and here our main goal will be to compare and contrast this spacetime with its two limits, namely the  
SV metric ($\gamma \rightarrow 1$) and the JNW metric ($c \rightarrow 0$) 
in terms of the effective potential encountered by a photon moving in these three spacetimes.\footnote{The observational 
	features of the above two cases are well studied in the literature for example see \cite{SV1} for SV metric and \cite{VK,sau} 
	for the JNW metric.} The standard approach to study the null geodesics motion and consequently finding the photon 
sphere, which are the unstable orbits of light rays, is to use  the Hamilton-Jacobi equation. 
After performing a standard analysis, the effective potential encountered by a photon in a spherically symmetric 
background can be written as \cite{ sau}
\begin{equation}\label{veff}
	V_{eff}(r)=g_{tt}\frac{C+L^2}{\mathcal{B}(r)}~,
\end{equation}
where $g_{tt}$ represents the temporal component of the metric and $\mathcal{B}(r)$ is its 
two sphere part. Here, $C$ and $L$ are the Carter's constant and the angular momentum respectively, and both of 
them are conserved quantities associated with the photon motion. Using the relevant functions 
from Eq. (\ref{RJNW}), we get the expression of the effective potential for the photon motion to be 
\begin{equation}
	V_{eff}(r)=\frac{C+L^2}{r^2+c^2} \Bigg(1-\frac{b}{\sqrt{r^2+c^2}}\Bigg)^{2\gamma-1}~.
\end{equation}
	
The circular photon orbits can be obtained by imposing the condition of vanishing radial velocity  
along with the condition for the extremum of the effective  potential, $\dot{r}=0$ and  $V^{\prime}_{eff}=0$. 
Typically this extremum turns out to be a maximum, and photon orbits are unstable. Solving this constraint, 
we obtain the following three solutions,
\begin{equation}\label{photonspheres}
	r_{1}=0,~~r_{2}=\sqrt{b^2-c^2},~~~\text{and}~~r_{3}=\Bigg\{b^2\Bigg(\frac{1}{4}\big(2\gamma -1\big)^2
	+\big(2\gamma -1\big)+1\Bigg) -c^2\Bigg\}^{1/2}=\sqrt{b^2\Big(\gamma+\frac{1}{2}\Big)^2-c^2}~.
\end{equation}
This gives correct result for SV metric ($\gamma \rightarrow 1$) \cite{SV1} and in the limit $c \rightarrow 0$, it 
correctly reduces to the known photon sphere of the JNW metric. Now, depending on the range of the various parameters 
$(b, c, \gamma)$, the above three roots give the photon sphere in this spacetime, as we elaborate below.
In this section we shall consider only values of $\gamma<1$.	
	
	\begin{figure}[h!]
		\begin{minipage}[b]{0.45\linewidth}
			\centering
			\includegraphics[width=0.6\textwidth]{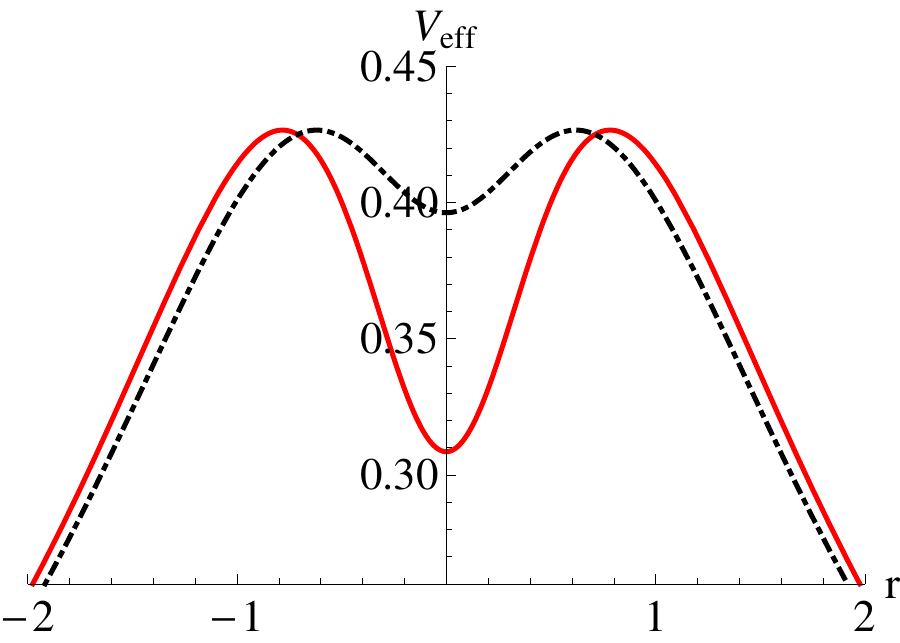}
			\caption{$V_{eff}$ for the WH branch when $c>b$. The solid red curve represents 
				the case with $c=1.1$, and the dotted black is with $c=1.2$. For both, $b=1, \gamma=0.85, C=L=1$.}
			\label{fig:veffc>b}
		\end{minipage}
		\hspace{0.2cm}
		\begin{minipage}[b]{0.45\linewidth}
			\centering
			\includegraphics[width=0.6\textwidth]{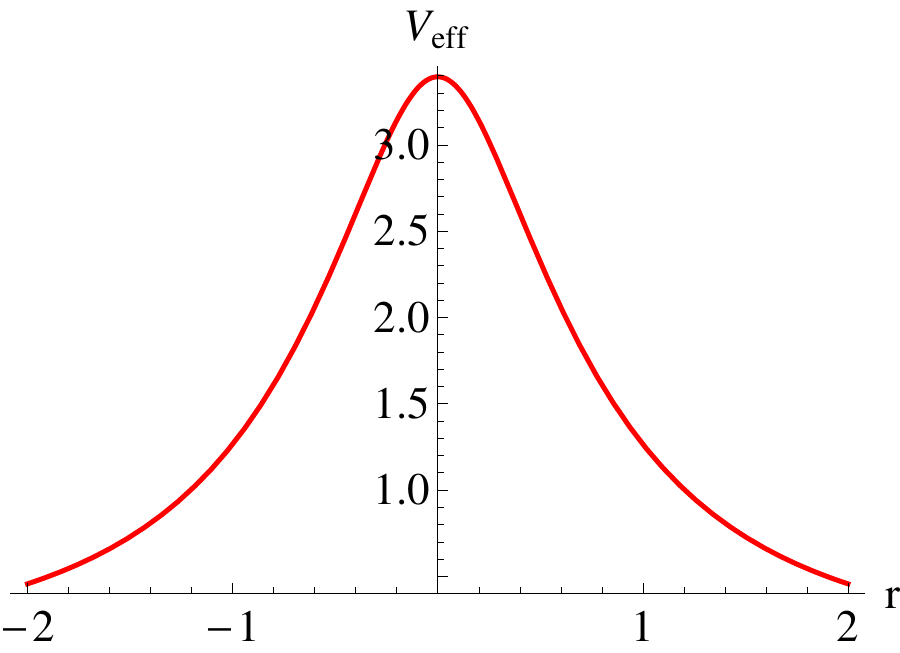}
			\caption{Plot of the effective potential for the WH case. The solid red curve represents 
				the case with $\gamma=0.35$ for which  maximum of $V_{eff}(r)$ is at $r_{1}$. We have set $b=1, c=1.1, C=L=1$.}
			\label{fig:Veffc>bgamma<0.5}
		\end{minipage}
	\end{figure}
	
	\noindent	
	$\bullet$~Photon motion for $b<c$ :
	As we have discussed above, in this case the spacetime is a wormhole (WH) with a throat at $r=0$. 
	We can directly see from Eq. (\ref{photonspheres}) that in this case, the solution $r_{2}$ is not real, 
	and the parameter range the other two solutions might be relevant here. In this context,  
	we note that the motion of light in wormhole backgrounds has been widely studied in the literature, 
	see, e.g., \cite{shaikh1} and references therein. Here the stand out feature for the wormhole  background  
	is the fact that apart from the conventional photon sphere outside the throat, the throat can itself work as 
	a location of the photon sphere (i.e., a location of the maximum of the effective potential), see \cite{shaikh2} for 
	details. For our purposes, the two sub-cases are important in this class of solutions, namely whether the 
	third root $r_3$ is real or not corresponding to some given parameter range.   
	
	First, we consider the case when $r_{3}$ is real. Then the root $r_3 $ ($>r_1$) can be real only if  
	$\gamma> \frac{1}{2}$. This range of gamma  $1/2<\gamma<1$ is the same as that of the JNW metric, and when this 
	condition is satisfied, $r_{3}$ is a location of photon sphere. The plot of the effective potential  
	is shown in fig. \ref{fig:veffc>b}, for two different values of $ c$ with fixed $\gamma=0.85, b=1$. 
	A light ray from our universe will encounter this photon sphere first and will form shadow \cite{shaikh3}. 
	Now the interesting case happens if we continue to increase $c$, such that the value of $r_{3}$ saturates.
	
	Next, consider the case where $r_{3}$ not real, which can be due to two different reasons. Firstly if  
	$\gamma< 1/2$, then $r_{3}$ is not physical for any  value of $b, c$ (with $b<c$). In this case, the throat at  
	$r_{1}$ acts as the location of the unstable orbits of photon and generates the shadow. We note that this is in
	contrast to JNW spacetime, where no photon sphere is possible for $\gamma< 0.5$. This case 
	is illustrated in fig. \ref{fig:Veffc>bgamma<0.5}.
	
\begin{figure}[h!]
		\begin{minipage}[b]{0.45\linewidth}
			\centering
			\includegraphics[width=0.6\textwidth]{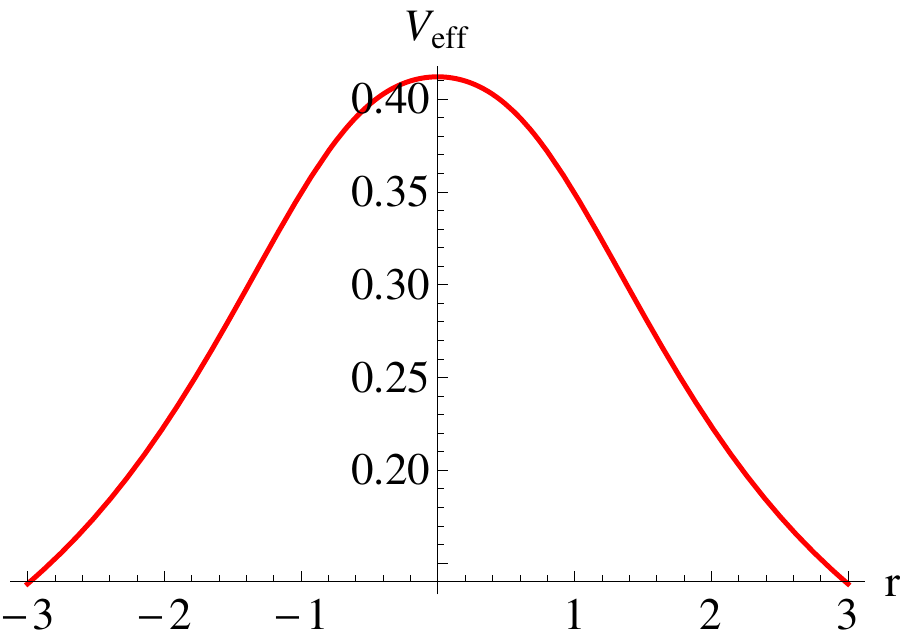}
			\caption{$V_{eff}$ for the WH case. The curve represents the case 
				with $\gamma=0.85$ for which  maximum of $V_{eff}(r)$ is at  $r_{1}$. Here, $b=1, c=1.5, C=L=1$.}
			\label{fig:veffc>>b}
		\end{minipage}
		\hspace{0.2cm}
		\begin{minipage}[b]{0.45\linewidth}
			\centering
			\includegraphics[width=0.6\textwidth]{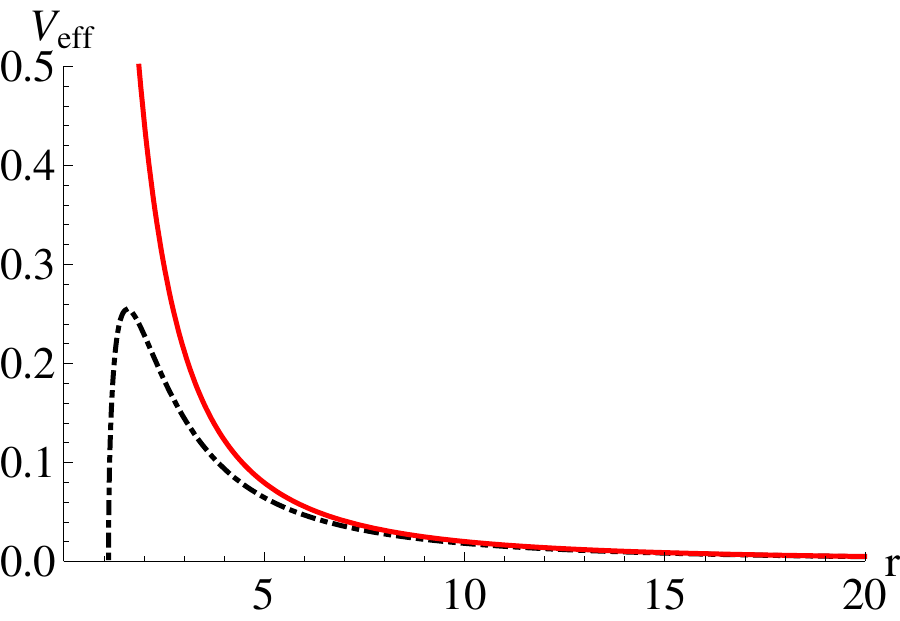}
			\caption{$V_{eff}$ for the NS case. The solid red curve is with 
				$\gamma=0.45$, and the dashed black curve is for $\gamma=0.85$. For both plots, $b=1.5, c= C=L=1$.}
			\label{fig:veffns}
		\end{minipage}
	\end{figure}

	On the other hand, even if  $\gamma> 1/2$, since $\gamma $ is always less than unity,  
	there is an upper bound on $c$ for which $r_{3}$ can be real. This can be found from setting 
	$\gamma=1$ in  $r_{3}$, from which we conclude that if $c>\frac{3b}{2}$,
	then $r_{3}$ is not real, and in this case also $r_{1}$ acts as a location of unstable photon orbit,
	see the plot of the  effective potential in Fig. \ref{fig:veffc>>b}.
	
	$\bullet$ ~ Photon motion for $b>c$ :
	In this case, the spacetime itself is valid upto a radial distance $r=\sqrt{b^2-c^2}$, 
	and it represents a naked singularity at that coordinate location. Clearly, among the three solutions in 
	Eq. (\ref{photonspheres}), the $r_1=0$ solution is not relevant here, and since the second 
	solution $r_{2}=\sqrt{b^2-c^2}$ is the location of the curvature singularity itself, it cannot be the physically 
	relevant photon sphere. Importantly however, depending on the range of $\gamma$, the solution $r_2$ decides 
	whether the third solution gives rise to a physically relevant photon sphere or not, since  
	if for some value of $\gamma$, it turns out that $r_{3}< r_{2}$, then $r_{3}$ is not physical, even if it a real solution. 
	
	The allowed range of $\gamma$ can be found from Eq.  (\ref{photonspheres}) to be $\gamma > \frac{1}{2}$, as we 
	are considering the case $b>c$. So in this case, the spacetime has \textbf{one photon sphere} at the coordinate 
	location $r_{3}$ for $\frac{1}{2}<\gamma<1$ and \textbf{no photon sphere} for $0<\gamma< \frac{1}{2}$. 
	Note that as long as $\gamma > \frac{1}{2}$, there is no restriction on the values of $b, c $ except $b>c$, 
	unlike above case, as $r_{3}$ will always be real as long as above conditions are satisfied. In Fig. \ref{fig:veffns},  
	we have shown the effective potential in this case. 
	
	Furthermore, unlike the WH case discussed above, here we cannot have a photon sphere for $\gamma < 0.5$, as $r_{3}< r_{2}$ 
	will always be satisfied, and we will have a NS configuration without a photon sphere. In this way the
	NS branch mimics the behaviour of photons in the JNW metric.   
	
	To summarize, the SV-JNW spacetime we have constructed can have one photon sphere (at $r_{3}$ or $ r_{1}$) or can 
	have no photon sphere at all, depending on the relative values of the parameters $b$ and $c$, as well as that of 
	$\gamma$. This behaviour is to be contrasted with the standard JNW metric and the SV metric, which are two particular 
	limits of our metric. To this end, a qualitative comparison of the effective potentials encountered by a massless particles 
	moving in SV, JNW and SV-JNW solutions is given in fig \ref{fig:Veffcomprison} 
	(for the SV and SV-JNW solutions, only the WH branches are plotted). In particular, note the contrast between the JNW and  
	SV-JNW cases. Due to the fact that the SV-JNW solution has a WH branch, its effective potential continues into the other 
	universe also. The large $r>0$ behaviours of all these cases are similar, since all of them are 
	asymptotically Minkowski spacetimes. 
	
\subsection{Energy conditions}

In this subsection, we shall analyze the energy conditions corresponding to the metric in Eq. (\ref{RJNW}). 
Since most of the wormhole spacetimes discussed in the literature violate the energy conditions, it is important 
to check this out for our metric as well.

The EM tensor can be calculated from the Einstein equations $G_{\mu\nu}=T_{\mu\nu}$. Since our 
considered geometry does not possess any event horizon, the components of the energy momentum tensor, given by 
$\rho=-T^{t}_{t}, p_{r}=T^{r}_{r},  p_{\theta}=T^{\theta}_{\theta}= T^{\phi}_{\phi}$ are valid for all 
values of the coordinate $r$ for which the metric is valid i.e., for the WH branch the range is $-\infty<r<\infty$, 
and given for the NS branch by $\sqrt{b^{2}-c^{2}}\leq r<\infty$. It is enough to check if the null energy condition (NEC) is violated
because if that is the case, then all the other energy conditions are violated as well. The NEC in this case is given 
from $\rho+p_{r} \geq 0$ and $\rho+p_{\theta} \geq 0$. Before proceeding, we recall that NEC is violated for 
SV spacetimes for all the values of the  parameter $c$, whereas the JNW metric satisfies the NEC everywhere. 	
	\begin{figure}[h!]
		\begin{minipage}[b]{0.45\linewidth}
			\centering
			\includegraphics[width=0.6\textwidth]{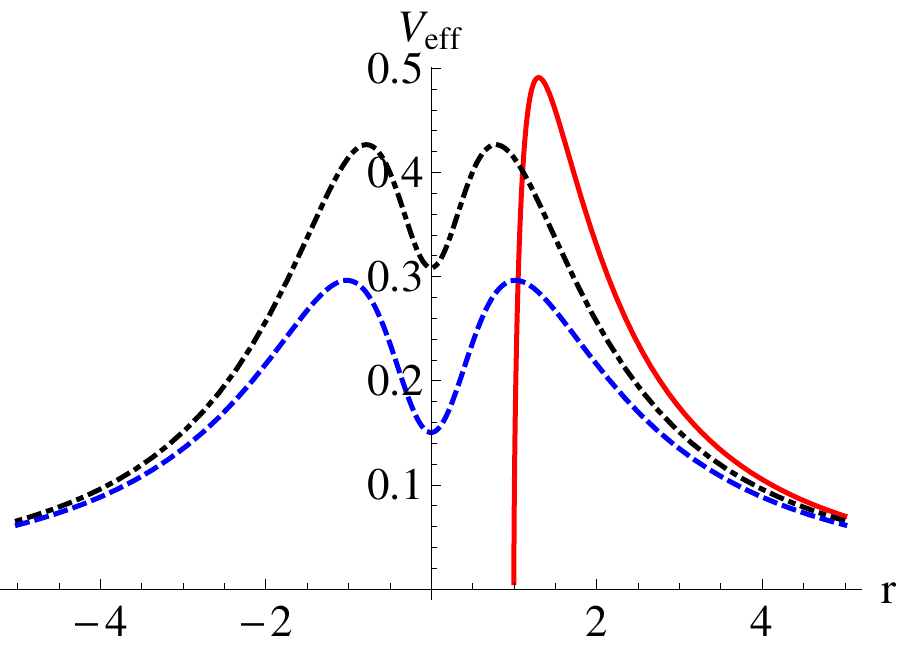}
			\caption{Effective potentials encountered by a massless particle for the SV (blue dashed curve), 
				JNW (red curve) and SV-JNW (black dot-dashed curve) spacetimes. }
			\label{fig:Veffcomprison}
		\end{minipage}
		\hspace{0.2cm}
		\begin{minipage}[b]{0.45\linewidth}
			\centering
			\includegraphics[width=0.6\textwidth]{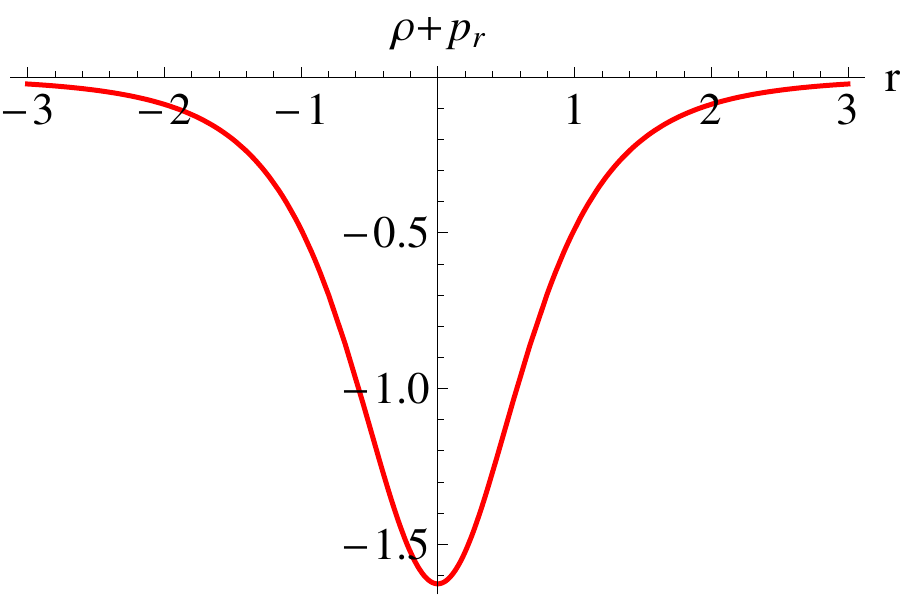}
			\caption{$\rho+p_{r}$  for the WH branch with $\gamma=0.45, c=1.1,b=0.1$. The NEC is always violated. }
			\label{fig:NEC1WH}
		\end{minipage}
	\end{figure}
	
\begin{figure}[h!]
		\begin{minipage}[b]{0.45\linewidth}
			\centering
			\includegraphics[width=0.6\textwidth]{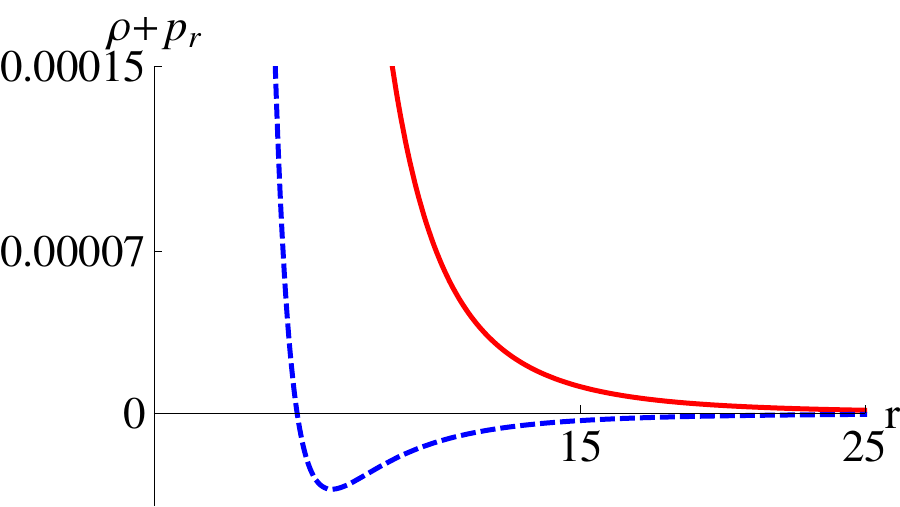}
			\caption{$\rho+p_{r}$ for the NS branch. The solid red curve is plotted with $\gamma=0.25$ which 
				satisfies the NEC, and the dashed blue
				curve is with $\gamma=0.85$  for which  the NEC is violated. Here, $b=1.4, c=0.5$.}
			\label{fig:NEC1NS}
		\end{minipage}
		\hspace{0.2cm}
		\begin{minipage}[b]{0.45\linewidth}
			\centering
			\includegraphics[width=0.6\textwidth]{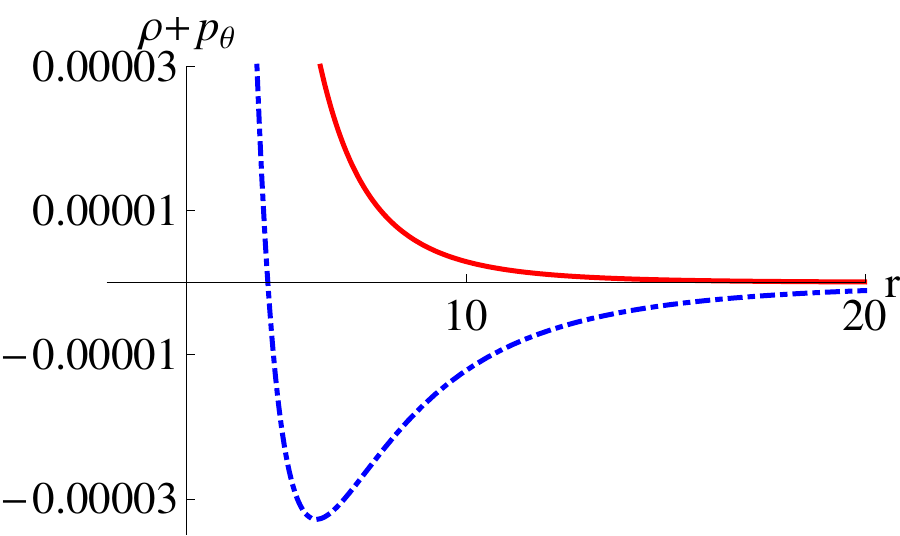}
			\caption{$\rho+p_{\theta}$ for the NS branch. The solid red curve is for
				$\gamma=0.25$ and the dashed blue curve is  for $\gamma=0.85$. Here, $b=1.4, c=0.5$. }
			\label{fig:NEC2NS}
		\end{minipage}
	\end{figure}

In Figures \ref{fig:NEC1WH} and \ref{fig:NEC1NS}, we have plotted $\rho+p_{r}$ for both the WH and NS branch of the 
metric of Eq. (\ref{RJNW}), with the parameters indicated in the respective captions. As can be seen, the WH branch 
violates the NEC everywhere, while for the NS branch, by decreasing the value of $\gamma$, this condition 
can be satisfied for all values of $r$. In this case only we need to check the second condition also. In 
Fig. \ref{fig:NEC2NS}, we have plotted the quantity $\rho+p_{\theta}$ for the same parameter values as in 
Fig. \ref{fig:NEC1NS}, and as can be seen, the plot with $\gamma=0.25$ satisfies the condition 
$\rho+p_{\theta}>0$ as well. Thus we conclude that, in contrast with the  SV solution, for sufficiently small 
values of $\gamma$, the NEC can be made to be satisfied here.	
	
It remains to check whether the solutions in the NS branch which satisfy the NEC (e.g. the red curves in 
Figs. \ref{fig:NEC1NS}, \ref{fig:NEC2NS}) satisfy other energy conditions such as the weak energy conditions and the
strong energy conditions as well. We have checked that these solutions in the NS branch satisfy the other 
energy conditions as well.
Thus in the NS branch of the SV-JNW  spacetime at least one set of solutions can be found which satisfies all the standard energy
conditions for some particular values of the parameters.

\subsection{Source of the SV-JNW solution}\label{Source_SVJNW}
An explanation of the above behavior of  the energy momentum tensor and hence the NEC can be given by the following
general considerations, as well as finding out the source of the SV-JNW metric following  \cite{BW, Bronnikov} (see also \cite{Canate:2022gpy}).
In this section we shall once again concentrate only on the branch $\gamma<1$.
We start by considering the following general static spherically symmetric line element 
\begin{equation}\label{static_metric}
	ds^2=-\mathcal{A}(r)\text{d}t^2+\mathcal{A}^{-1}(r)\text{d}r^2+\mathcal{B}^2(r)\text{d}\Omega^2~.
\end{equation}
Assuming that this metric satisfies the Einstein equations of the form $G^\mu_\nu=T^{\mu}_{\nu}$ we obtain, 
\begin{equation}\label{rho+pr}
	G^{t}_{t}-G^{x}_{x}=2\mathcal{A}(r)\frac{\mathcal{B}^{\prime\prime}(r)}{\mathcal{B}(r)}=-(\rho+p_r)~.
\end{equation}
Since for the branch of the SV-JNW spacetime we are considering, $\mathcal{A}>0$ for ranges of $r$  for which the metric is 
valid, the sign of second derivative of the areal radius determines whether the condition $(\rho+p_r)\geq 0$ 
is satisfied or not.  For our solution we calculate the required quantity to be 
\begin{equation}
	\frac{\mathcal{B}^{\prime\prime}(r)}{\mathcal{B}(r)}=\frac{4c^2(c^2+ r^2)+b^2 (2 c^2 + r^2 (-1 + \gamma)) 
		(1 + \gamma)-2 b c^2 \sqrt{c^2 + r^2}(3+\gamma)}{4 (c^2 + r^2)^2 (b - \sqrt{c^2 + r^2})^2}~.
\end{equation}
Now it can be easily seen that with $\gamma<1$, depending on the relative values of the parameters $b$ and $c$
this quantity can be positive or negative. In particular for the WH branch with $c>b$ this quantity is everywhere
positive for the entire respective  range of $r$, while for the NS branch with $b>c$ the sign can be both positive or negative 
depending on the value of $\gamma$. With  given values of  $b> c$, for relatively higher values of $\gamma$, a range of 
$r$ coordinate can be  found where this quantity is positive  - thereby violating the NEC.

These observations can be more clearly understood by finding out the source of the SV-JNW spacetime.
In the context of the SV regularization of the Schwarzschild solution as it was recently 
shown in \cite{BW},
a combination of a scalar field  $\phi(r)$ and a nonlinear electromagnetic field, both minimally coupled to
gravity, can act as a source of such spacetimes (see also \cite{Canate:2022gpy}). 
Furthermore, in \cite{Bronnikov} it was 
also established that the scalar and nonlinear electromagnetic field can act as the source of an
arbitrary static spherically symmetric metric.  In this section we  find out the scalar field and the 
nonlinear electrodynamic (NED) field corresponding to the  SV-JNW solution.	
	
We consider the following matter action  
\begin{equation}
	\mathcal{S}_m=\int\sqrt{-g}d^4x\bigg[-\frac{1}{2}h(\phi)\partial_\mu\phi\partial^\mu\phi-V(\phi)
	+\mathcal{L}(\mathcal{F})\bigg]~,
\end{equation}
which contains an action of a scalar field $\phi(r)$ and that of a  nonlinear electromagnetic field $\mathcal{L}(\mathcal{F})$.
Here $h(\phi)$ is a function of $\phi$ and $V(\phi)$ is the corresponding potential. 
The Lagrangian density of the NED field $\mathcal{L}(\mathcal{F})$ is written
as a function of $\mathcal{F}=F_{\mu\nu}F^{\mu\nu}$, 
where the electromagnetic field tensor $F_{\mu\nu}$ is defined in terms
of the vector potential in the usual way $F_{\mu\nu}=\partial_\mu A_{\nu}-\partial_\nu A_{\mu}$.

The energy-momentum tensor obtained by varying the above action contains two parts corresponding to the 
scalar field and the NED respectively : $T_{\mu}^{\nu}=\Phi_{\mu}^{\nu}[\phi]+N_{\mu}^{\nu}[F]$.
Explicitly, these are given by 
\begin{equation}\label{EM_phi_and F}
	\begin{split}
		\Phi_{\mu}^{\nu}[\phi]=\frac{1}{2}h(\phi)\partial_\mu\phi\partial^\nu\phi-
		\delta_\mu^\nu\bigg[\frac{1}{4}h(\phi)\partial_\sigma\phi\partial^\sigma\phi+\frac{V(\phi)}{2}\bigg]~,\\
		N_{\mu}^{\nu}[F]=2\mathcal{L}_\mathcal{F}F_{\mu\sigma}F^{\mu\sigma}-\frac{1}{2}\delta_\mu^\nu\mathcal{L}(\mathcal{F})~.
	\end{split}
\end{equation}
Here, we have used the notation $\mathcal{L}_\mathcal{F}=\frac{d \mathcal{L}(\mathcal{F})}{d\mathcal{F}}$.	
	
Now assuming that $\phi=\phi(r)$ and the source of the electromagnetic field is a magnetic monopole of charge
$q$, i.e. $F_{\theta\phi}=q\sin \theta$, the spherical symmetry of the system restricts the components of the above EM tensor
to be 
\begin{equation}\label{EM_Tensor}
	\begin{split}
		\Phi_{\mu}^{\nu}(r)=\frac{1}{4}h(r)\mathcal{A}(r)\phi^{\prime 2}(r) ~\text{diag}\big(-1,1,-1,-1\big)
		-\delta_\mu^\nu\frac{V}{2}~,\\
		N_{\mu}^{\nu}(r)=-\frac{1}{2}~ \text{diag}\big(\mathcal{L},\mathcal{L},\mathcal{L}-2\mathcal{F}\mathcal{L}_{\mathcal{F}},
		\mathcal{L}-2\mathcal{F}\mathcal{L}_{\mathcal{F}}\big)~.
	\end{split}
\end{equation}
Here the prime denotes a derivative with respect to $r$. Notice that here the invariant is given by 
$\mathcal{F}=2q^2/\mathcal{B}^4(r)$. Using the Einstein equation $G^{nu}_\mu=T^{\nu}_\mu$
we obtain the following relations between the metric components and the components of the EM tensor
\begin{equation}
	\begin{split}\label{Einstein_eqs}
		G^{t}_t-G^{r}_r=T^{t}_t-T^{r}_r~ \rightarrow ~ \frac{\mathcal{B^{\prime\prime}}(r)}{\mathcal{B}(r)}
		=-\frac{1}{4}h(r)\phi^{\prime 2}(r)~,\\
		G^{t}_t-G^{\theta}_\theta=T^{t}_t-T^{\theta}_\theta=-\mathcal{F}\mathcal{L}_{\mathcal{F}}~,
	\end{split}
\end{equation}
where we have used Eq. (\ref{rho+pr}) in the second expression of the first equation. 
We now use these equations to obtain the required scalar field and the nonlinear electromagnetic
Lagrangian density.	
	
\begin{figure}[h!]
		\begin{minipage}[b]{0.45\linewidth}
			\centering
			\includegraphics[width=0.6\textwidth]{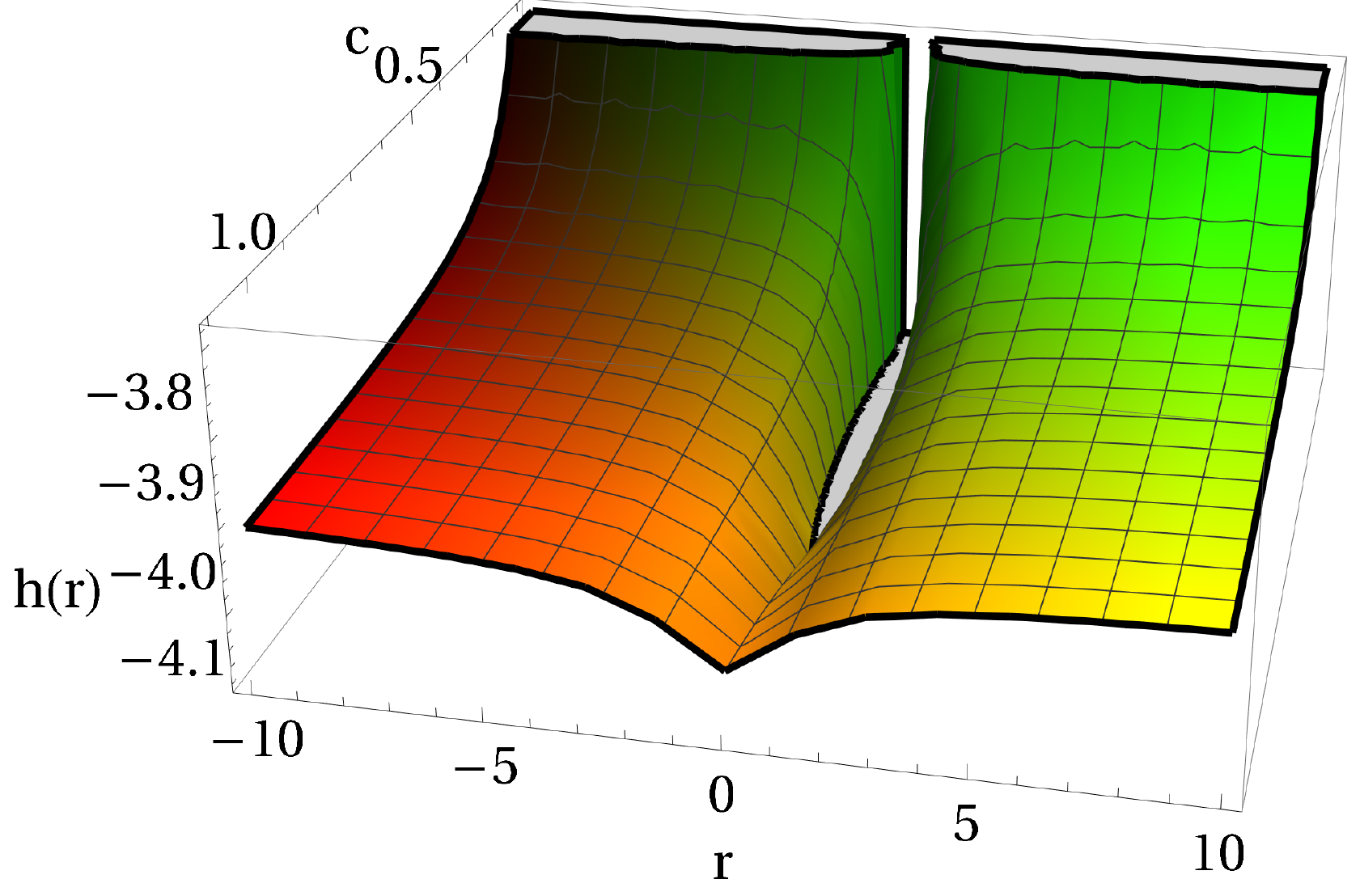}
			\caption{Plot of the the function $h(r)$ which determines the sign of the kinetic energy of the scalar field
				in the WH branch. Here $\gamma=0.45$ and $b=0.1$.  }
			\label{fig:KE_WH}
		\end{minipage}
		\hspace{0.2cm}
		\begin{minipage}[b]{0.45\linewidth}
			\centering
			\includegraphics[width=0.6\textwidth]{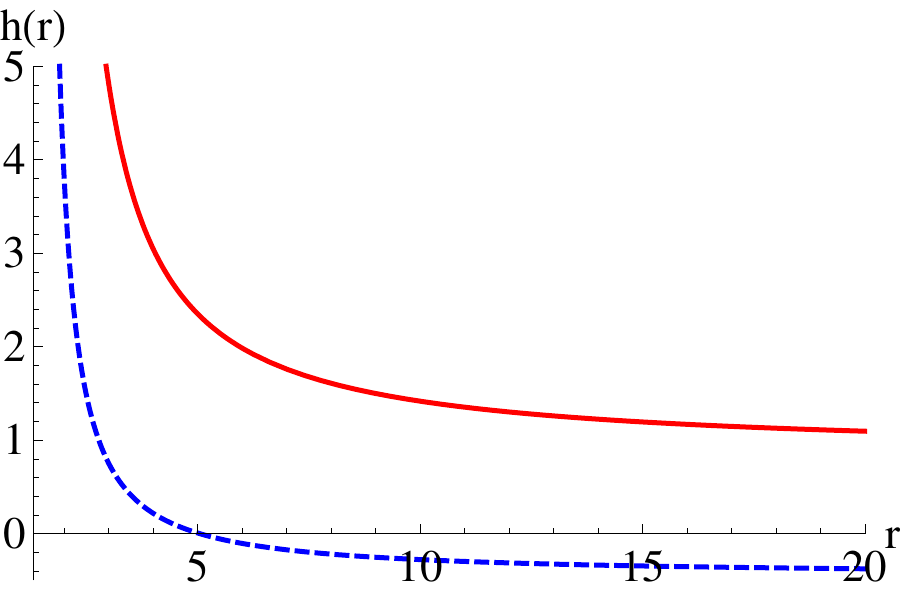}
			\caption{Plot of the the function $h(r)$ in the NS branch. The red curve is with $\gamma=0.25$ and
				the blue curve is for $\gamma=0.85$. For both the plots $b=1.4, c=0.5$. }
			\label{fig:KE_NS}
		\end{minipage}
	\end{figure}

First we obtain the function $h(r)$. Using the prametrization freedom of the 
scalar field we choose $\phi(r)$ to be (this choice has been used in \cite{BW,Bronnikov}), 
\begin{equation}\label{scalar_r}
	\phi(r)=\arctan (r/c)~.
\end{equation}
Substituting this form of the scalar field in the first equation of (\ref{Einstein_eqs}) along with the
form of the areal radius $\mathcal{B}(r)$ for the metric in Eq. (\ref{RJNW}) we can obtain the functional 
dependence of $h(r)$. However the expression for it is lengthy and we do not provide it here, rather 
we will be interested the limiting behaviour of this functions at a few particular coordinate
locations, namely $r \rightarrow 0$ and $r \rightarrow \pm \infty$.   These are respectively given
by
\begin{equation}
	h(r)~\Big|_{r \rightarrow 0}= -\frac{2(b-2c+b \gamma)}{b-c}~, \quad \text{and} \quad 
	h(r)~\Big|_{r \rightarrow \pm \infty}=-\frac{4c^2+b^2(\gamma^2-1)}{c^2}~.
\end{equation}
Now it is easy to see that for the WH branch $b<c$, the function $h(r)$ is  negative at the throat,
indicating that the scalar field is phantom there. 
Furthermore, for this branch this function always remains negative
for the entire range of the coordinate $r$, indicating that the scalar field is phantom in nature
for the entire coordinate range of the   WH branch. 
This is in contrast with the results obtained in \cite{Bronnikov}, where a variation of the SV-JNW solution 
was constructed and it was shown that the function determining the sign of the kinetic energy 
can change sign depending on the value of the SV regularization parameter $c$. In Fig. \ref{fig:KE_WH},
we have plotted the function $h(r)$ in the WH branch. From which it is easy to see that for the entire 
range of $r$, the function $h(r)$ does not change sign and is always negative.	
	
On the other hand in the NS branch ($b>c$), similar to the behavior of the energy conditions, the nature of the scalar field 
depends on the parameter $\gamma$. In Fig. \ref{fig:KE_NS} we have plotted $h(r)$ for two values $\gamma$
with fixed values of $b$ and $c$ in the NS branch. As can be easily seen for higher value of $\gamma$ 
this function changes sign and hence the scalar field becomes  phantom from canonical, whereas for a
relatively lower value of $\gamma$ the scalar field is always canonical.	
	
Next, we can find out the NED Lagrangian density $\mathcal{L}(\mathcal{F}(r))$ as a solution of the following
first order differential equation obtained from the second equation of (\ref{Einstein_eqs}):
\begin{equation}
	\frac{d \mathcal{L}}{dr}=\frac{bc^2 r (2\gamma+1) \Big(1-\frac{b}{\sqrt{c^2+r^2}}\Big)^\gamma \Big(b(\gamma+1)-2\sqrt{c^2+r^2}\Big)}
	{\big(c^2+r^2\big)^{3}\Big(b-\sqrt{c^2+r^2}\Big)^{2}}~.
\end{equation}
The solution for this equation is a smooth and positive  function of $r$  in both the WH and the NS branches of the SV-JNW solution.
The explicit expression for $\mathcal{L}$ is cumbersome and we do not provide it here. Rather we note here that unlike the
case of the (analogous) SV regularized Fisher solution presented in \cite{Bronnikov} this NED Lagrangian density does not reduce to 
zero for some specific value of $\gamma$. For the solution presented in \cite{Bronnikov} the density $\mathcal{L}$ is zero for
$\gamma=1/2$, so that  source of the metric is  only the scalar field.

\section{The deformed JMN metric}

As a second example of using the SV method to regularise naked singularities, we now study the SV modified version of the JMN metric,
which is an equilibrium end state geometry of a realistic gravitational collapse model \cite{JMN}. For a 
specific profile of energy density and tangential pressure, the spherically symmetric JMN metric is given by 
\begin{equation}\label{JMN}
	ds^2=-\big(1-M_{0}\big)\bigg(\frac{r}{r_{b}}\bigg)^{\frac{M_{0}}{1-M_{0}}}\text{d}t^2 + \frac{\text{d}r^2}{1-M_{0}}+ r^2 \text{d}\Omega^2~.
\end{equation} 
Here $r_{b}$ is a matching radius, and $M_{0}$  is related to the Schwarzschild mass, to which this metric 
can be matched smoothly. This metric represents an unstable equilibrium configuration of a collapsing star 
supported by a non zero tangential pressure and vanishing radial pressure. 

As discussed in \cite{JMN},  if we confine our attention to the parameter region $0<M_{0}<1$, then the metric 
does not possess any event horizon, and there is a curvature singularity at $r=0$. Thus, the range of the 
parameter $M$, to which we shall restrict ourselves here, is  $0<M_{0}<1$. It was also shown in the same 
work that in this geometry, there is at least one geodesic that can reach to an asymptotic observer that 
terminates at the singularity in the past, which points to the fact that the metric given in Eq. (\ref{JMN})  
is in fact a naked singularity. Our motivation in this section would be to check that whether 
this central singularity can removed by the SV modification, and study the resulting geometry. To this end, 
we first write down the resulting metric using the procedure of \cite{franzin2}, which is given by, 
\begin{equation}\label{RJMN}
	ds^2=-\big(1-M_{0}\big)\bigg(\frac{\sqrt{r^2+c^2}}{r_{b}}\bigg)^{\frac{M_{0}}{1-M_{0}}}\text{d}t^2 + 
	\frac{\text{d}r^2}{1-M_{0}}+ \big(r^2+c^2\big) \text{d}\Omega^2~.
\end{equation} 
Here as usual, $c$ is a real positive quantity. The range of the radial coordinate is $r= -\infty$ to $r =\infty$, 
and the range of other coordinates remain same. We shall call this metric the SV-JMN solution. Now a little 
inspection shows that this metric actually is a traversable wormhole with a throat at $r=0$ for all values of 
the parameter $c$. To see this, we first note that, as mentioned before, for $M_{0}<1$, the metric does not possess 
an event horizon. Secondly, the `flaring out' condition of a traversable wormhole can be checked from considering 
the area of the two sphere 
$A(r)=(r^2+c^2)$, which should satisfy the conditions $ A^{\prime}(r=r_{0})=0$, 
and $A^{\prime\prime} (r)=2 >0$. Hence the throat is the position of the minimum area, and the wormhole is two way traversable. 
It can also be checked that the Ricci scalar goes as $R\sim(r^2+c^2)^{-2}$, and is finite everywhere including  $r=0$. 
Thus we conclude that the geometry in  Eq. (\ref{RJMN}) is indeed a wormhole.	
	
\subsection{Photon motion in SV-JMN background}
Now we briefly discuss the motion of massive and massless particles in the SV-JMN geometry, 
and compare them with those in the original JMN metric, the latter being studied in detail in \cite{JMN},\cite{shaikh4}. 
Using the standard formula for the motion of photons in spherically symmetric spacetimes given in Eq. (\ref{veff}), 
we find the effective potential to be
\begin{equation}
	V_{eff}(r)=\big(1-M_{0}\big)\Bigg(\frac{\sqrt{r^2+c^2}}{r_{b}}\Bigg)^\frac{M_{0}}{1-M_{0}}\Bigg(\frac{C+L^2}{r^2+c^2}\Bigg)~,
\end{equation}
and its derivative with respect to $r$ is given by
\begin{equation}
	V^\prime_{eff}(r)=r\Big(r^2+c^2\Big)^\frac{5M_{0}-4}{2(1-M_{0})}~~.
\end{equation}
Clearly, we have only one physically relevant solution for the photon sphere, namely $r=0$, 
which is just the location of the throat of the  WH geometry. The corresponding effective 
potential looks like the standard effective potential for the wormhole metric, when the throat acts 
as a position of the photon sphere - an example of which is already given in Fig. \ref{fig:veffc>>b}. 
	
The components of the energy momentum tensor calculated from the metric in Eq. (\ref{RJMN}) are given by
\begin{equation}
	\rho=\frac{M_{0}r^{2}+c^{2}\big(2M_{0}-1\big)}{\big(c^{2}+r^{2}\big)^{2}}~~,~~p_{r}=-\frac{c^{2}}
	{\big(c^{2}+r^{2}\big)^{2}}~~,~~\text{and}~~p_{\theta}=p_{\phi}=-\frac{M_{0}^{2}r^{2}+2c^{2}
		\Big(M_{0}^{2}-3M_{0}+2\Big)}{4\big(c^{2}+r^{2}\big)^{2}\big(M_{0}-1\big)}~.
\end{equation}
As can be seen directly from these expressions, the components are regular at $r\rightarrow 0$ and reduces to 
the JMN values in the limit $c\rightarrow 0$ (the role of negative pressure in gravitational collapse leading
to naked singularities was investigated in \cite{Cooperstock}). In contrast to the original JMN spacetime, there is a non zero radial pressure 
here, which  is always negative and vanishes at $c\rightarrow 0$ as it should. Since the original JMN metric 
represents the unstable equilibrium configuration of a gravitationally collapsing star with only the tangential components of 
pressure being non zero, it will be interesting to study whether our metric also represents an equilibrium configuration 
of a collapsing matter cloud  with non zero radial pressure as well as a tangential pressure. We leave this for the future. 

To check if the energy conditions are violated, it is enough in this case to check that NEC is violated for all values of 
the parameters. This can be readily seen from the expressions of $\rho$ and $p_{r}$ given above : since $M_{0}<1$, the 
quantity $\rho+p_{r}$ is always negative for sufficiently small values of $r$, and hence the NEC is violated.	
	
\subsection{Source of the SV-JMN}
Next we shall find out the possible source of the SV-JMN solution following the method discussed in \ref{Source_SVJNW}.
First we notice that the line element of Eq. (\ref{RJMN}) written in the $(t,r,\theta,\phi)$ 
coordinates is  not of the form in Eq. (\ref{static_metric}). It is possible to make a coordinate
transformation such that we can cast it as that of Eq. (\ref{static_metric}), however the resulting 
coordinates expressions are difficult manipulate analytically. Rather we  work with the line element 
of Eq. (\ref{RJMN}) and modify the results of section \ref{Source_SVJNW} suitably.

Here we need to write down the components of the EM tensor obtained from Eq. (\ref{EM_phi_and F})
for a general metric of the form
\begin{equation}\label{generic_metric}
	ds^2=-\mathcal{A}_1(r)\text{d}t^2+\mathcal{A}_2^{-1}(r)\text{d}r^2+\mathcal{B}^2(r)\text{d}\Omega^2~.
\end{equation}
It can be easily checked that the EM tensor corresponding to the scalar part in Eq. (\ref{EM_Tensor})
changes to 
\begin{equation}
	\Phi_{\mu}^{\nu}(r)=\frac{1}{4}h(r)\mathcal{A}_2(r)\phi^{\prime 2}(r) ~\text{diag}\big(-1,1,-1,-1\big)
	-\delta_\mu^\nu\frac{V}{2}~,
\end{equation}
Using the same procedure outlined in section \ref{Source_SVJNW} and using Eq. (\ref{scalar_r}) as the
choice of the scalar field we obtain the expression for the function $h(\phi)$ to be
\begin{equation}
	h(\phi)=\frac{2 M_0 \tan \phi ^2}{1-M_0}-4~=\frac{2 M_0 r ^2}{c^2(1-M_0)}-4~.
\end{equation}
Close to the throat, $h(r)$ is always negative, indicating that the scalar field is phantom
near the throat. On the other hand at $r \rightarrow \pm \infty$ this diverges, so that the 
for some intermediate value of $r$ the scalar field changes its nature. 	
	\begin{figure}[h!]
		\begin{minipage}[b]{0.45\linewidth}
			\centering
			\includegraphics[width=0.6\textwidth]{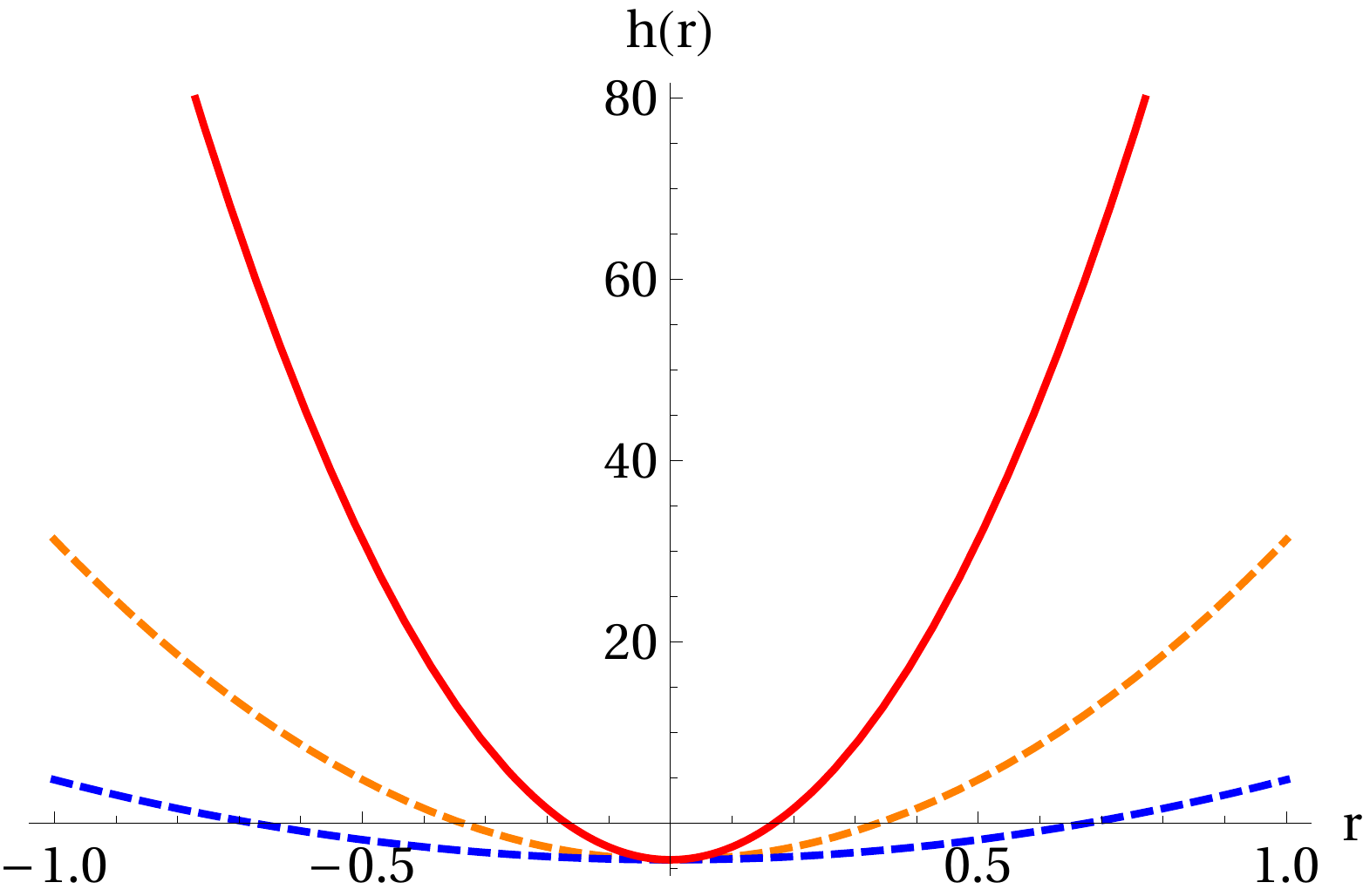}
			\caption{Plot of the function $h(r)$ for the regular JMN metric. Here $M=0.15$,
				and $c=0.05$ (red), $c=0.1$ (orange), and $c=0.2$ (blue) respectively. }
			\label{fig:scalar_JMN}
		\end{minipage}
		\hspace{0.2cm}
		\begin{minipage}[b]{0.45\linewidth}
			\centering
			\includegraphics[width=0.6\textwidth]{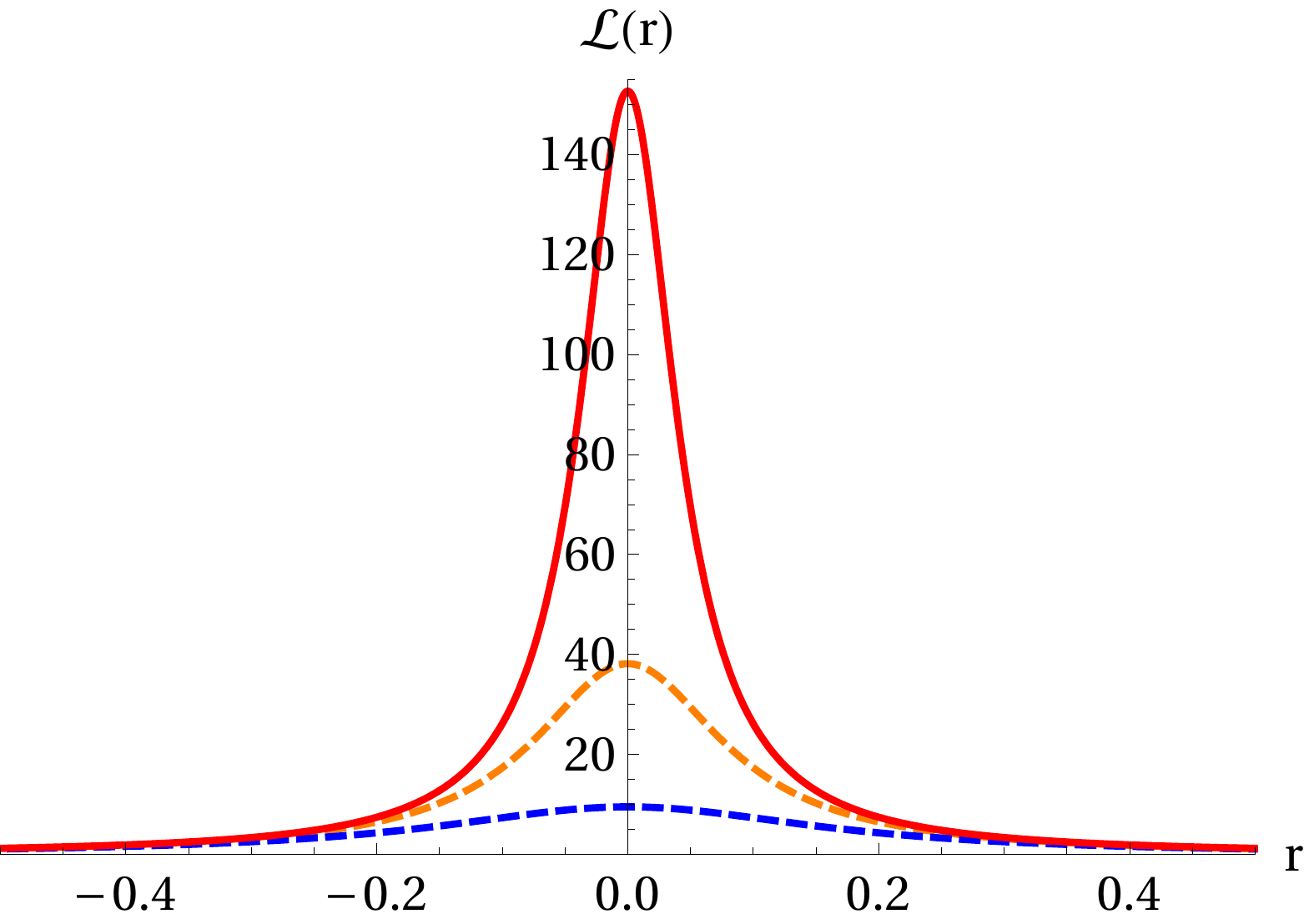}
			\caption{Plot of the NED Lagrangian density $\mathcal{L}(\mathcal{F}(r))$ for the regular 
				JMN metric. Here $M=0.15$,
				and $c=0.05$ (red), $c=0.1$ (orange), and $c=0.2$ (blue) respectively. }
			\label{fig:ned_JMN}
		\end{minipage}
	\end{figure}
Plots of the function with respect to $r$ for different values of $c$ are shown in 
Fig. \ref{fig:scalar_JMN}. As can be seen, this function is negative very close to the WH throat at $r=0$
for any values of the constant $c$. However by decreasing the constant $c$ sufficiently, the negative
part of the function $h(r)$ can be confined in a very small region around the throat.	   
	
In this case we can also provide  the analytic expression for the NED Lagrangian density. Written
as a function of $r$ this is given by
\begin{equation}
	\mathcal{L}(r)=\frac{M_0 \Big(c^2(9M_0-10)+2(3M_0-4)r^2\Big)}{4(c^2+r^2)^2(M_0-1)}~.
\end{equation}
This function is plotted in Fig. \ref{fig:ned_JMN} for different values $c$. We see that this is 
maximum near the throat and goes to zero at $r \rightarrow \pm \infty$.	
	
Finally, we point out that the SV-JMN metric we have constructed above is valid for all values of $r$, 
namely, $-\infty<r<\infty$. However, the JMN configuration was constructed to be the interior of collapsing matter 
distribution, matched with an exterior Schwarzschild solution of mass $M_{sch}=M_{0}r_{b}/2$, 
where the matching radius $r=r_{b}$. It may be worth mentioning that by redefining the constant $r_{b}$ as $\sqrt{r_{b}^{2}+c^{2}}$,
the metric in Eq. (\ref{RJMN}) can be matched with an exterior SV metric of the form
\begin{equation}\label{SV}
	ds^2=-\Bigg(1-\frac{M_{0}\sqrt{r_{b}^2+c^2}}{\sqrt{r^2+c^2}}\Bigg) \text{d}t^2 + 
	\Bigg(1-\frac{M_{0}\sqrt{r_{b}^2+c^2}}{\sqrt{r^2+c^2}}\Bigg)^{-1} \text{d}r^2 + 
	\Big(r^2+c^2\Big) \text{d}\Omega^2~.
\end{equation}
Here $M_{sv}=M_{0}\sqrt{r_{b}^2+c^2}/2$ with $M_{sv}$ being the mass appearing in the SV metric. 
Obviously the constant  $r_{b}$ appearing here is different from the one in Eq. (\ref{RJMN}), and we also  
note that for a given $M_{sch}=M_{sv}$, the values of $r_{b}$ are different for JMN and SV-JMN metrics. 
In particular, for a given value of $M_{sch}=M_{sv}=M$, the SV-JMN solution has smaller matching radius 
than its JMN cousin. Since the metric outside is SV (rather than Schwarzschild), we expect the properties of accretion  
disks corresponding to this model to be different from those discussed in \cite{JMN} and this might be 
interesting to investigate. Furthermore, it will be interesting to study the collapse of realistic matter with 
SV taken as the exterior to see whether SV-JMN solution discussed above arises as an equilibrium configuration. 
We hope to return to this problem in a future work.	
	
\section{Quasinormal modes of the regularised JNW metric}

Having investigated properties of the regularised JNW metric in section \ref{JNWsection}, we now address the important question of the stability of the JNW metric, focusing on the physically interesting wormhole branch (represented by the parameter ranges $b<c$ and $\gamma<1$). Specifically, in this section we calculate the quasinormal modes (QNMs) of the wormhole branch of the solution. As is well known, quasinormal modes are the eigenmodes of perturbations of spacetime. In general, QNMs are complex values, where the real parts give the oscillation frequencies, while the imaginary parts signify the damping rates. Hence, for a dissipative solution like a black hole or wormhole, positivity of the imaginary part of the eigenmodes would indicate instability. Quasinormal modes have long been used to prove stability of black hole spacetimes (see, for example, \cite{Chandrasekhar:1975zza, Iyer:1986np, Iyer:1986nq}) and recently have been used to investigate stability of wormholes as well in several works \cite{Kim:2008zzj, Churilova:2020aca, Churilova:2021tgn, DuttaRoy:2022ytr}. 	
	
We begin by considering a scalar perturbation in the  regularised JNW background, whose form is given by,
\begin{equation}
	\frac{1}{\sqrt{-g}} \partial_{\mu}\Big(\sqrt{-g} g^{\mu \nu} \partial_{\nu} \Psi(t,r, \theta ,\phi)\Big)=0~.
\end{equation}
Using the explicit form of the metric in Eq. (\ref{generic_metric}), this equation can be written as (with $\mathcal{A}_1(r)=\mathcal{A}_2(r)=f(r)$) 
\begin{multline}\label{fulleq}
	\frac{\partial_{t}^{2} \Psi}{f(r)}+  \frac{1}{\mathcal{B}^2(r)}\left(\partial_{r} (\mathcal{B}^2(r)) f(r) \partial_{r} \Psi + \mathcal{B}^2(r) f^{\prime}(r) \partial_{r} \Psi+\mathcal{B}^2(r) f(r) \partial_{r}^{2} \Psi\right)   \\
	+\frac{1}{\mathcal{B}^2(r)}\left(\frac{1}{\sin \theta} \partial_{\theta} \sin \theta \partial_{\theta} \Psi+\frac{1}{\sin ^{2} \theta} \partial_{\phi}^{2} \Psi\right)=0~.
\end{multline}	
	
	Since the regular JNW metric above is spherically symmetric, the scalar field can be assumed to be 
	of the form
	\begin{equation}\label{decomposition}
		\Psi (t,r, \theta ,\phi)=\frac{\psi(t,r)}{\mathcal{B}(r)} Y_{l m}(\theta, \phi)~,
	\end{equation}
	where $Y_{l m}(\theta, \phi)$ represent spherical harmonic function, and $\psi(t,r)$ is the function of radial coordinate $t,r$. Substituting Eq. (\ref{decomposition}) in Eq. (\ref{fulleq}),  after a separation of variables  we arrive at 
	\begin{equation}\label{sepratedeq}
		\frac{\partial^{2}\psi(r, t)}{\partial t^{2}}-\frac{\partial^{2}\psi(r, t)}{\partial r_{*}^{2}}+V(r)\psi(r, t) =0~,
	\end{equation}
	here $r_{*}$ is the tortoise coordinate, which is defined as,
	\begin{equation}
		dr_{*}=\frac{1}{f(r)} dr~,
	\end{equation}
	and $\ell$ is the separation constant for the angular parts. If we assume a harmonic time dependence $\psi(r,t)=u(r)e^{-i\omega t}$, Eq. (\ref{sepratedeq}) becomes
	\begin{equation}\label{scheq}
		\frac{d^{2}u(r)}{d r_{*}^{2}}+\left( \omega^{2}-V_{\text{scalar}}(r) \right) u(r) =0~,
	\end{equation}  with the exact form of the effective potential written as 
	\begin{equation}
		V_{\text{scalar}}(r)=\frac{f(r)l(l+1)}{\mathcal{B}^2(r)}+\frac{f^{2}(r)\partial_{r}^2 \mathcal{B}(r)}{\mathcal{B}(r)}+\frac{f(r)\partial_{r} f(r)\partial_{r}\mathcal{B}(r)}{\mathcal{B}(r)}~.
\end{equation}
	The final expression  of  Eq. (\ref{scheq}) has the form of a one dimensional Schrodinger equation with potential $V_{\text{scalar}}$. 
	Here, we note that in a similar manner the effective potential for electromagnetic ($s=1$) and gravitational ($s=2$) 
	perturbtations can be expressed as,
	\begin{equation}
		V_s(r)=f(r)\Bigg[\frac{l(l+1)}{\mathcal{B}^2(r)}+\frac{(1-s^2)}{\mathcal{B}(r)}
		\Big(f(r)\partial_{r}^2 \mathcal{B}(r)
		+\partial_{r} f \partial_{r} \mathcal{B}(r)\Big)\Bigg]~.
	\end{equation}
	Note that in the above equation, putting $s=0$ would result in the effective potential for scalar perturbation. 
	Henceforth, we shall mainly concentrate our analysis on the scalar field perturbation, since results for the 
	electromagnetic and gravitational cases are not qualitatively different.
	
	\begin{figure}[htp]
		{
			\begin{tabular}{cc}
				\includegraphics[width = 0.5\textwidth]{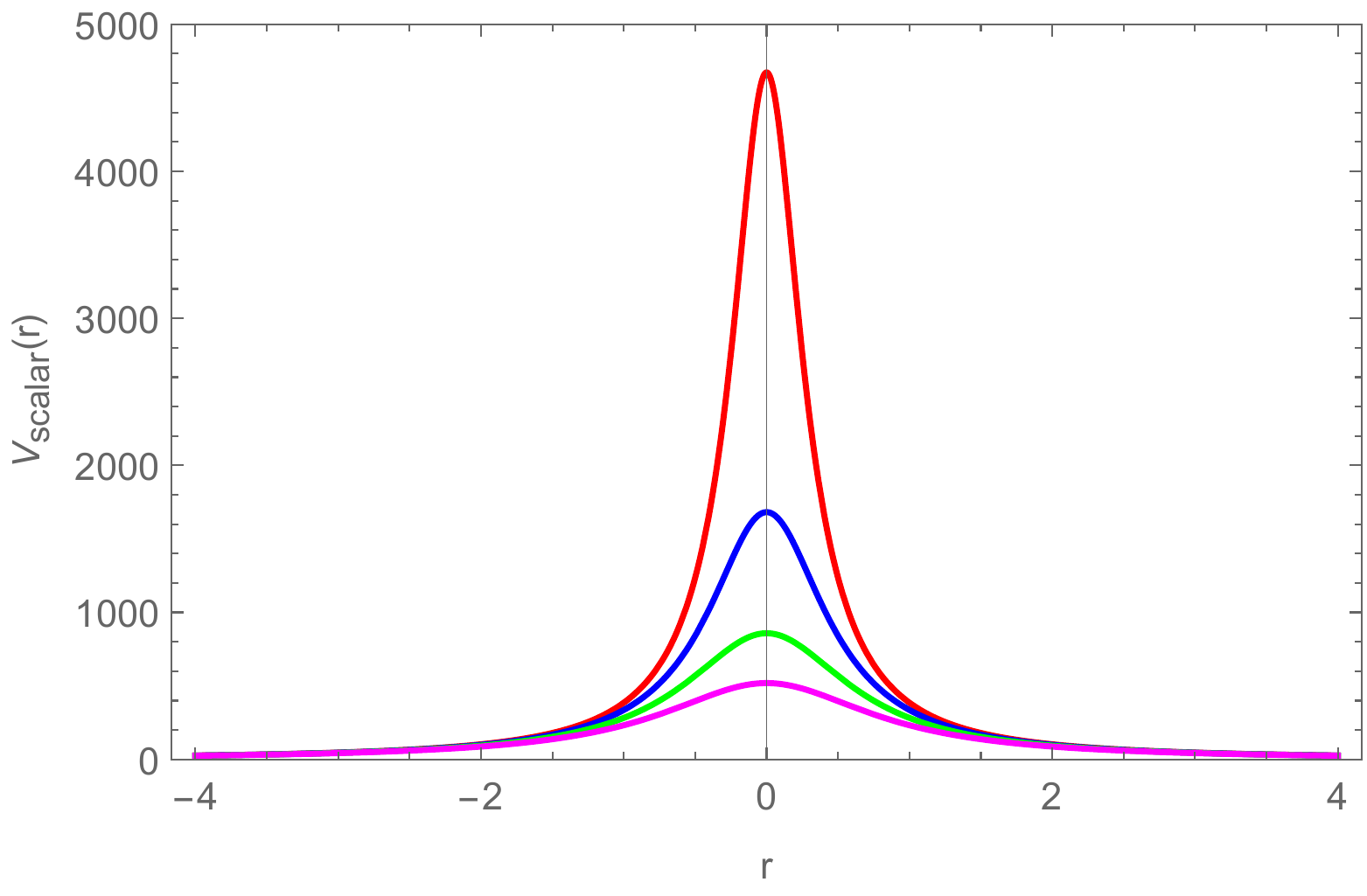} &
				\includegraphics[width = 0.5\textwidth]{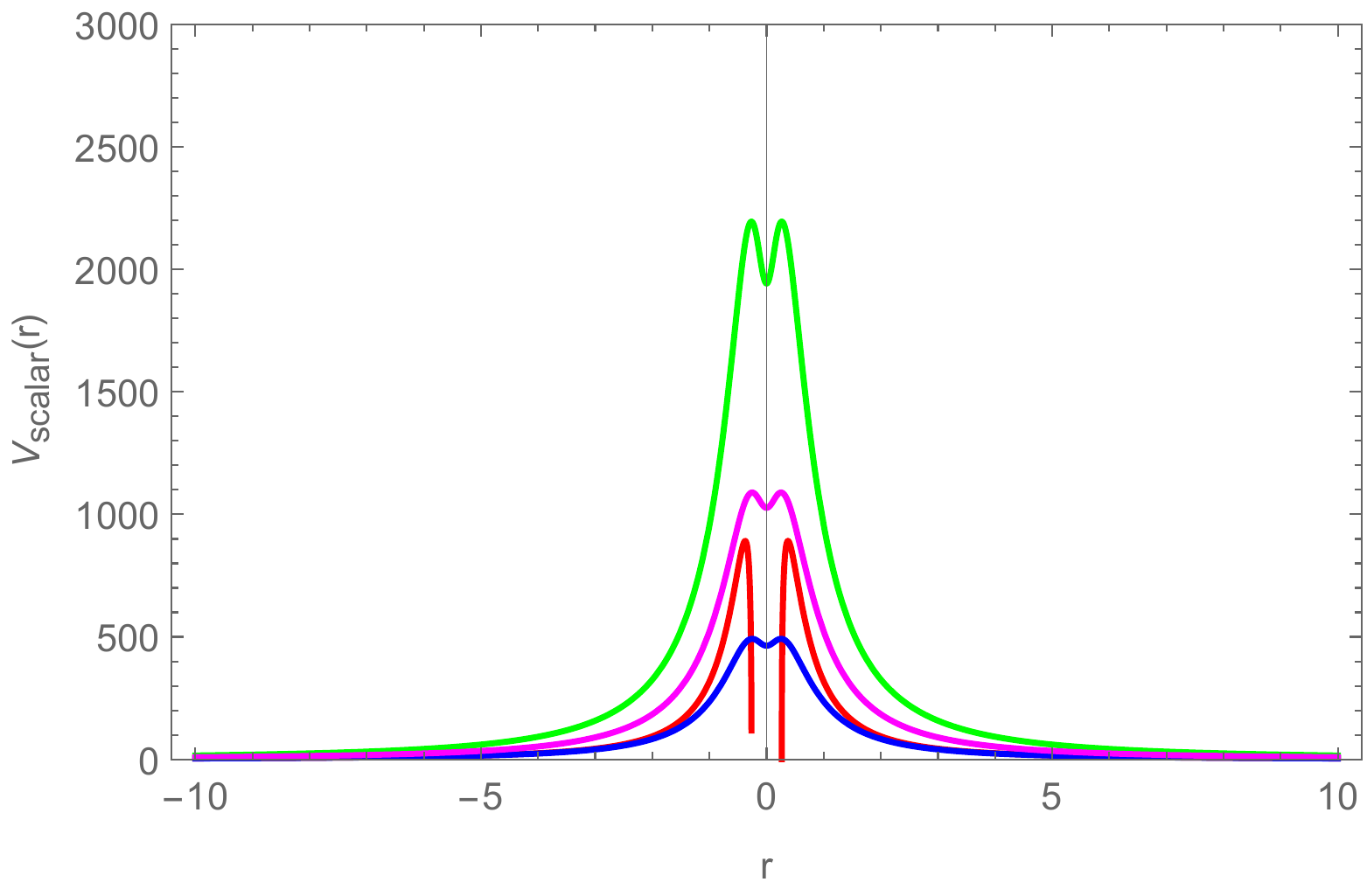}\\
				\includegraphics[width = 0.5\textwidth]{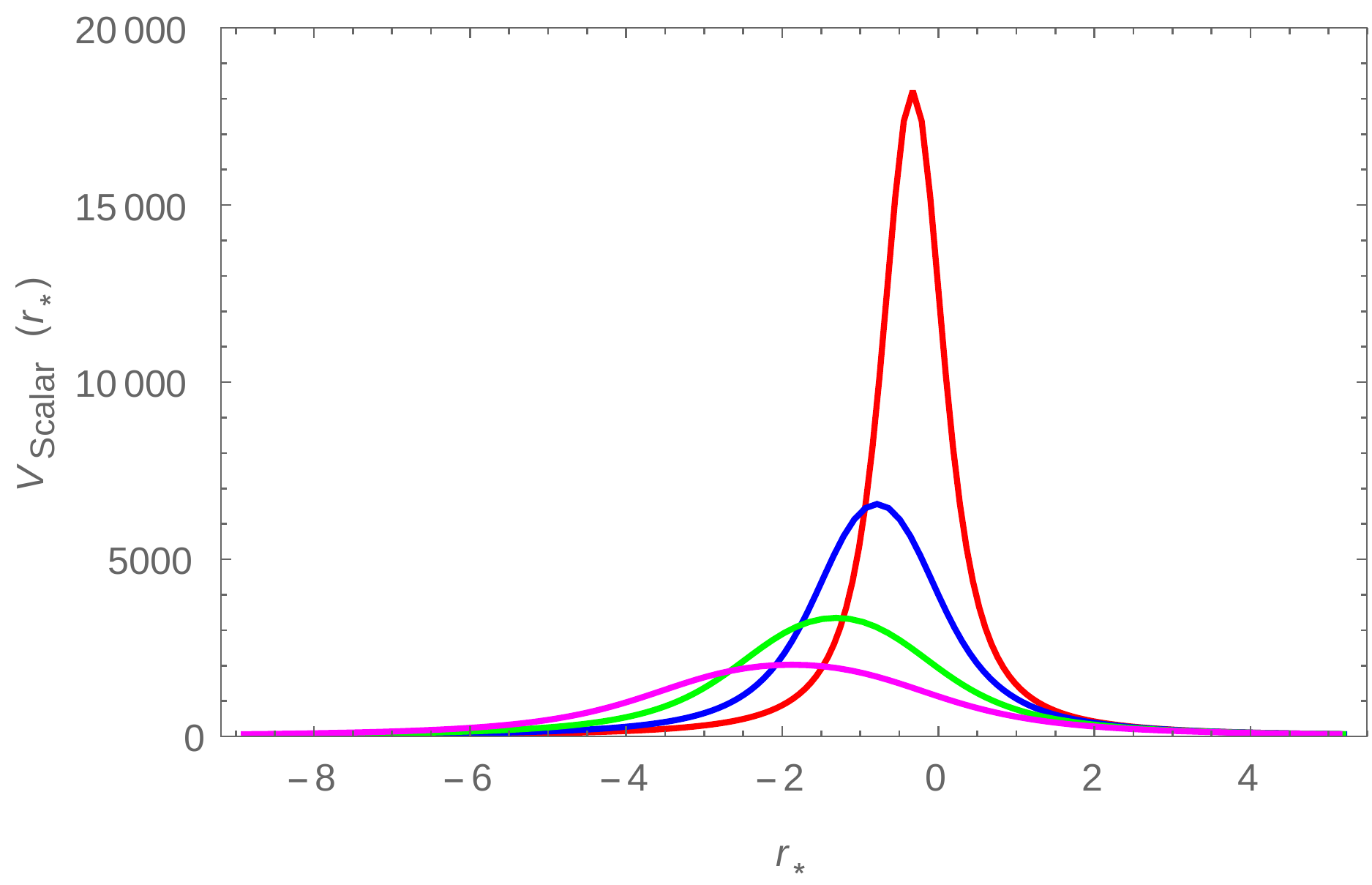} &
				\includegraphics[width = 0.5\textwidth]{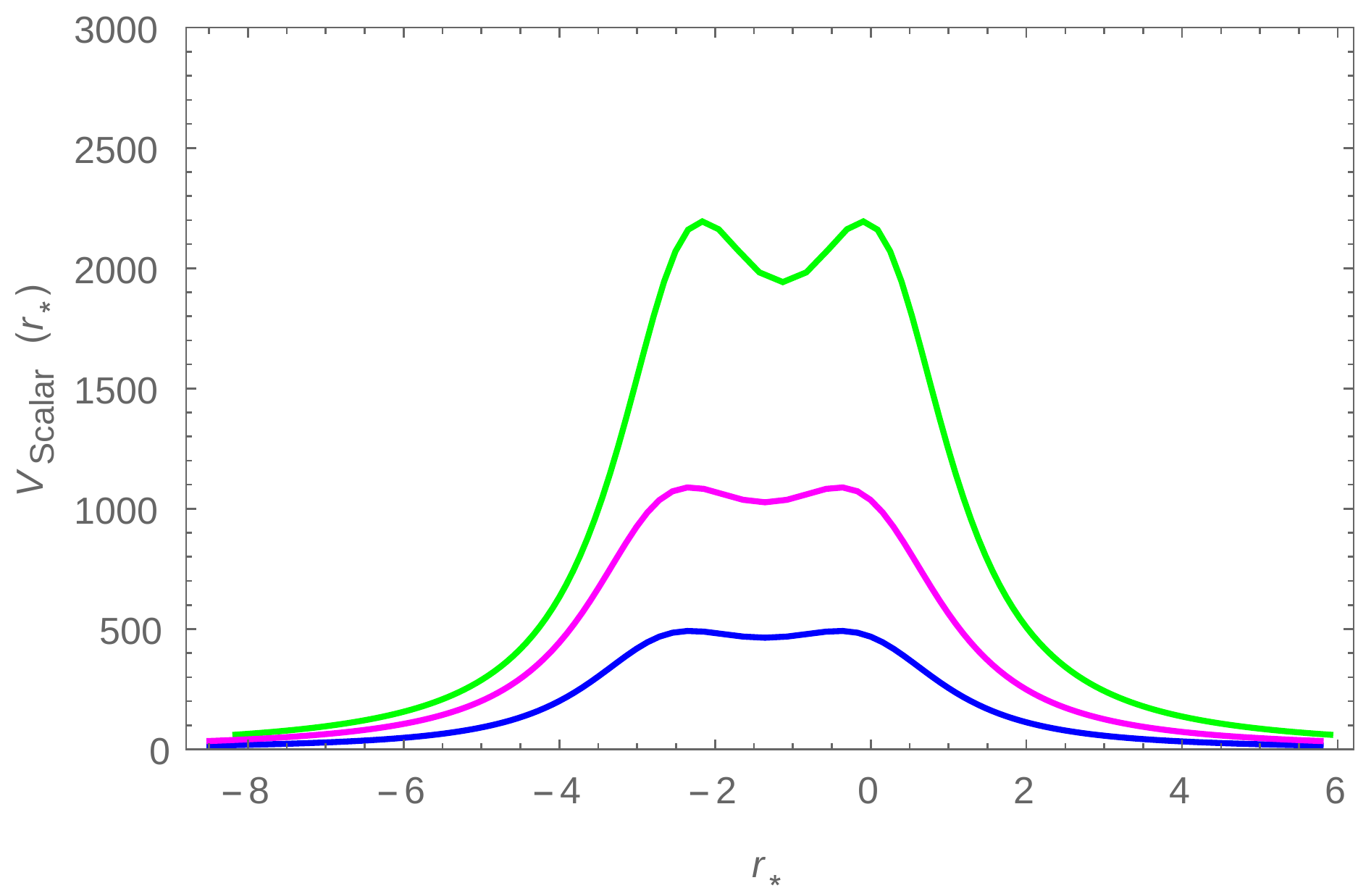} \\
				(a) & (b) \\
				(c) & (d) 
			\end{tabular}
			\caption{Variation of the potential $V_{scalar}(r)$ against $r$ and $r_{*}$. The red, blue, green and magenta curves denote $(b, c, \gamma, \ell) = (0.2, 0.3, 0.5, 20), (0.4, 0.5, 0.5, 20), (0.6, 0.7, 0.5, 20), (0.8, 0.9, 0.5, 20)$ respectively in panel (a) and $(0.4, 0.3, 0.7, 20), (0.4, 0.5, 0.9, 20), (0.3, 0.4, 1.1, 40), (0.4, 0.5, 0.9, 30)$ in panel (b). The parameter choices for the red curve in panel (b) ensure that it denotes the potential for the naked singularity solution. In panels (c) and (d), the same colour coding is followed as in (a) and (b) respectively. Note that in panel (d), the curve corresponding to the naked singularity is not given because of the difficulty in performing the numerical integration for $r_{*}(r)$.}
			\label{fig:Vrvsr_rs}}
	\end{figure}
	
	The plot for the effective potential $V_{\text{scalar}}$ of the scalar field is given in Fig. \ref{fig:Vrvsr_rs} for various values of the parameters $b$, $c$, $\gamma$ and $\ell$. Panels (a) and (b) show the variation of $V_{\text{scalar}}$ versus $r$, whereas in panels (c) and (d), $V_{\text{scalar}}$ has been plotted against $r_{*}$, which has been obtained by numerical integration. Panels (a) and (c) of Fig. \ref{fig:Vrvsr_rs} show that for several values of the parameters (all of which ensure that we are choosing the wormhole branch of the solution), the effective potential exhibits a single peak, with two turning points. Panels (b) and (d) show some parameter choices where a double peak is exhibited. This is similar to other wormhole solutions in the literature \cite{DuttaRoy:2022ytr}. The figure also shows that $V(r, r_{*} \rightarrow \infty) \approx 0$ for the wormhole branch. From the wave equation, it is clear that asymptotically the solution takes the form $\Psi \approx e^{\pm i \omega t}$. This ensures that the physically relevant solutions are purely outgoing and incoming at infinity and the wormhole throat respectively. \footnote{It is well known that the throat of the WH plays the same role as that of a BH event horizon in case of absence of any reflecting potential at  $r_{*}\rightarrow -\infty$.} The form of the effective potential suggests the use of the WKB method for calculating the QNMs of the wormhole. At the same time, panel (b) of Fig. \ref{fig:Vrvsr_rs} shows multiple peaks for the effective potential for several other parameter choices.   However, it is important to note that this method would not be valid for calculating QNMs in case of the naked singularity because of the markedly different structure of the effective potential. Since the WKB method has been used extensively in the literature for calculating QNMs (see, for example \cite{Iyer:1986np, Iyer:1986nq}), we shall not describe it here. Briefly, the expression for the QNMs obtained by this method is, 
	\begin{equation}
		\omega^{2}=\left[ V_{0}+\sqrt{-2V^{\prime\prime}
			_{0}}\Lambda\left(  n\right)  -i\left(  n+\frac{1}{2}\right)  \sqrt{-2V^{\prime\prime}
			_{0}}\left(  1+\Omega\left(  n\right)  \right)  \right]~,
	\end{equation}
	with
	\begin{equation}
		\Lambda\left(  n\right)  =\frac{1}{\sqrt{-2V^{\prime\prime}
				_{0}}}\left[  \frac{1}{8}\left(  \frac{V_{0}^{\left(  4\right)  }}{V_{0}^{\prime\prime}
		}\right)  \left(  \frac{1}{4}+\alpha^{2}\right)  -\frac{1}{288}\left(
		\frac{V_{0}^{\prime\prime\prime}
		}{V_{0}^{\prime\prime}
		}\right)  ^{2}\left(  4+60\alpha^{2}\right)  \right]~, 
	\end{equation}
	and 
	\begin{multline}
		\Omega\left(  n\right)  =\frac{1}{-2V_{0}^{\prime\prime}}\left[  \frac{5}{6912}
		\left( \frac{V_{0}^{\prime\prime\prime}}{V_{0}^{\prime\prime
		}}\right)  ^{4}\left(  77+188\alpha^{2}\right)  -\frac{1}{384}
		\left(
		\frac{V_{0}^{\prime\prime\prime 2}V_{0}^{\left(  4\right)  }}{V_{0}^{\prime\prime3}}\right)  
		\left(  51+100\alpha^{2}\right)  +\right.  \\
		\left.  \frac{1}{2304}\left(  \frac{V_{0}^{\left(  4\right)  }}{V_{0}^{\prime\prime}}\right) ^{2}
		\left(  67+68\alpha^{2}\right)  +\frac{1}{288}\left(
		\frac{V_{0}^{\prime\prime\prime}V_{0}^{\left(  5\right)  }}{V_{0}^{\prime\prime 2}}\right)  
		\left(  19+28\alpha^{2}\right)  -\frac{1}{288}\left(  \frac
		{V_{0}^{\left(  6\right)  }}{V_{0}^{\prime\prime
		}}\right)  \left(  5+4\alpha^{2}\right)  \right]  .
	\end{multline}
	Here, $\alpha$ and  $V_{0}^n$ are given by
	\begin{equation}
		\alpha=n+\frac{1}{2}~, ~~~V_{0}^n=\frac{d^nV}{dr_{*}^n}\Big|_{r_{*}=r_{*}(r_{max})}~,
	\end{equation}
	with $n=0, 1, 2.....$. In the above equation, the prime denotes differentiation with respect to $r_{*}$, which can be translated to differentiation with respect to $r$ by use of the chain rule. Also, $V_0 \equiv V_{scalar}(r_0)$, where $r_{*} = r_0$ denotes the extremum of the potential.
	
\begin{figure}[htp]{
			\begin{tabular}{cc}
				\includegraphics[width=0.5\textwidth]{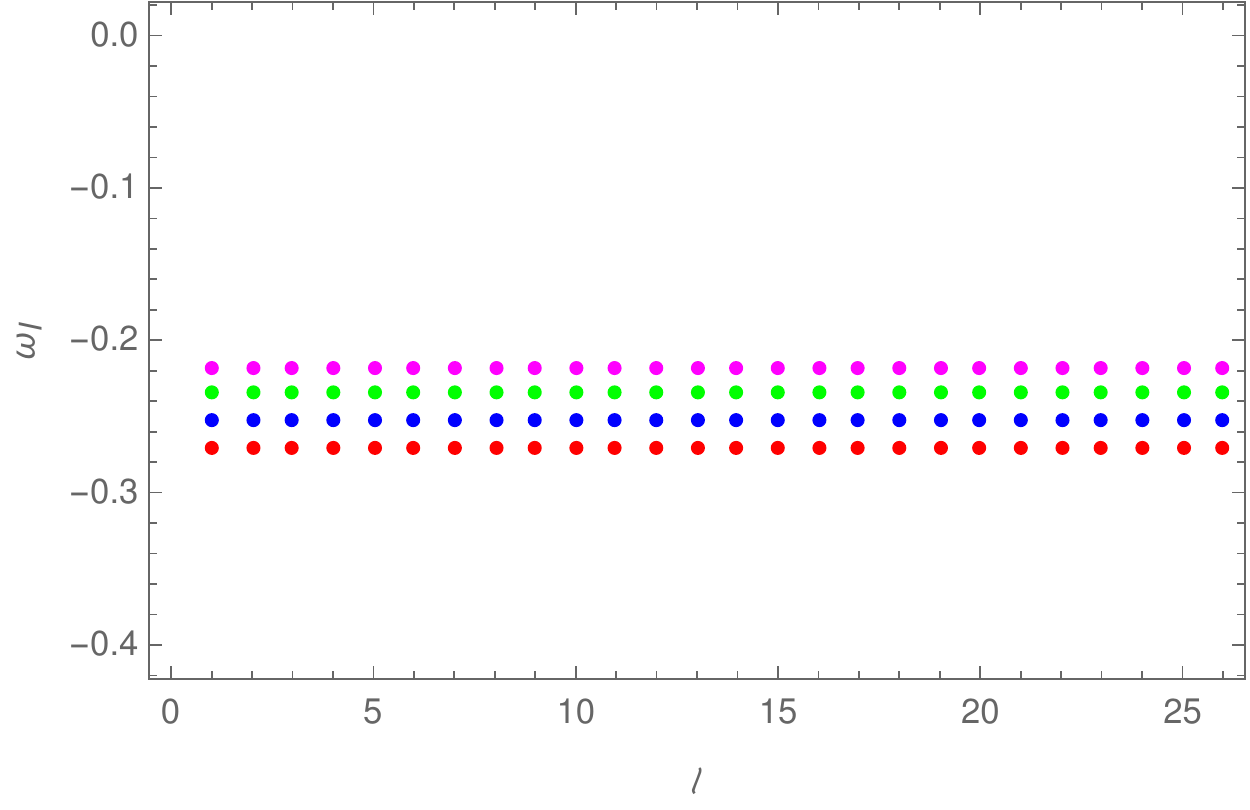}&
				\includegraphics[width=0.5\textwidth]{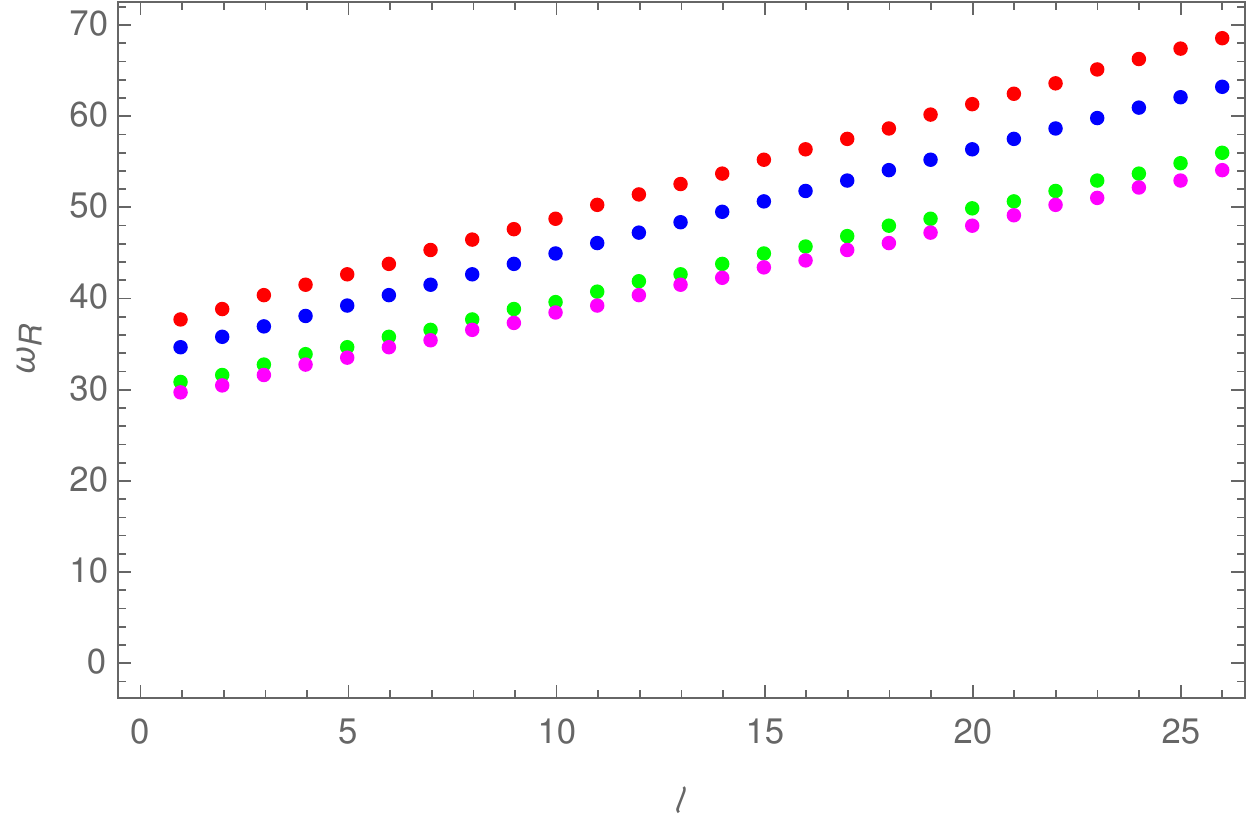}\\
				(a) & (b)
			\end{tabular}
			\caption{Variation of real and imaginary parts of $\omega$ ($n=0$) against $\ell$. Here, $b=0.4$, $c=0.5$ and $\gamma =0.2, 0.3, 0.4, 0.5$ for the red, blue, green and magenta curves respectively.}
			\label{fig:wvsl}}
	\end{figure}

	We shall primarily be interested in the fundamental $n=0$ frequency. Before employing the WKB method to calculate QNMs, it is necessary to ensure that, for the parameter values chosen, the effective potential exhibits only a single peak, thus ensuring two turning points. For example, this would include Fig. \ref{fig:Vrvsr_rs}(a) and (c), and would rule out the parameter choices corresponding to Fig. \ref{fig:Vrvsr_rs}(b) and (d). It is instructive to plot the real and imaginary parts of $\omega$ against $\ell$. This is shown in Fig. \ref{fig:wvsl}(a) and (b), from which it is clear that the higher values of $l$ do not necessarily dominate the spectrum of QNMs due to the magnitude of the imaginary component of the QNMs not appreciably varying as $\ell$ is changed. 
	
	It is also interesting to study the effect of the variation of the parameters $b$ and $c$ on the QNMs. This is physically interesting because it may be recalled from section \ref{JNWsection} that the parameter regime $b > c$ would ensure that the solution would be a naked singularity instead of a wormhole. Hence, by keeping $c$ constant and varying $b$ such that $b < c$, it would be possible to check whether the QNMs indicate any change in the wormhole solution as we get closer to the naked singularity branch. \footnote{It is important to reiterate that the QNMs of the naked singularity branch cannot be calculated by the WKB method (see Fig. \ref{fig:Vrvsr_rs}(b)).}
	
	\begin{figure}[htp]{
			\begin{tabular}{cc}
				\includegraphics[width=0.5\textwidth]{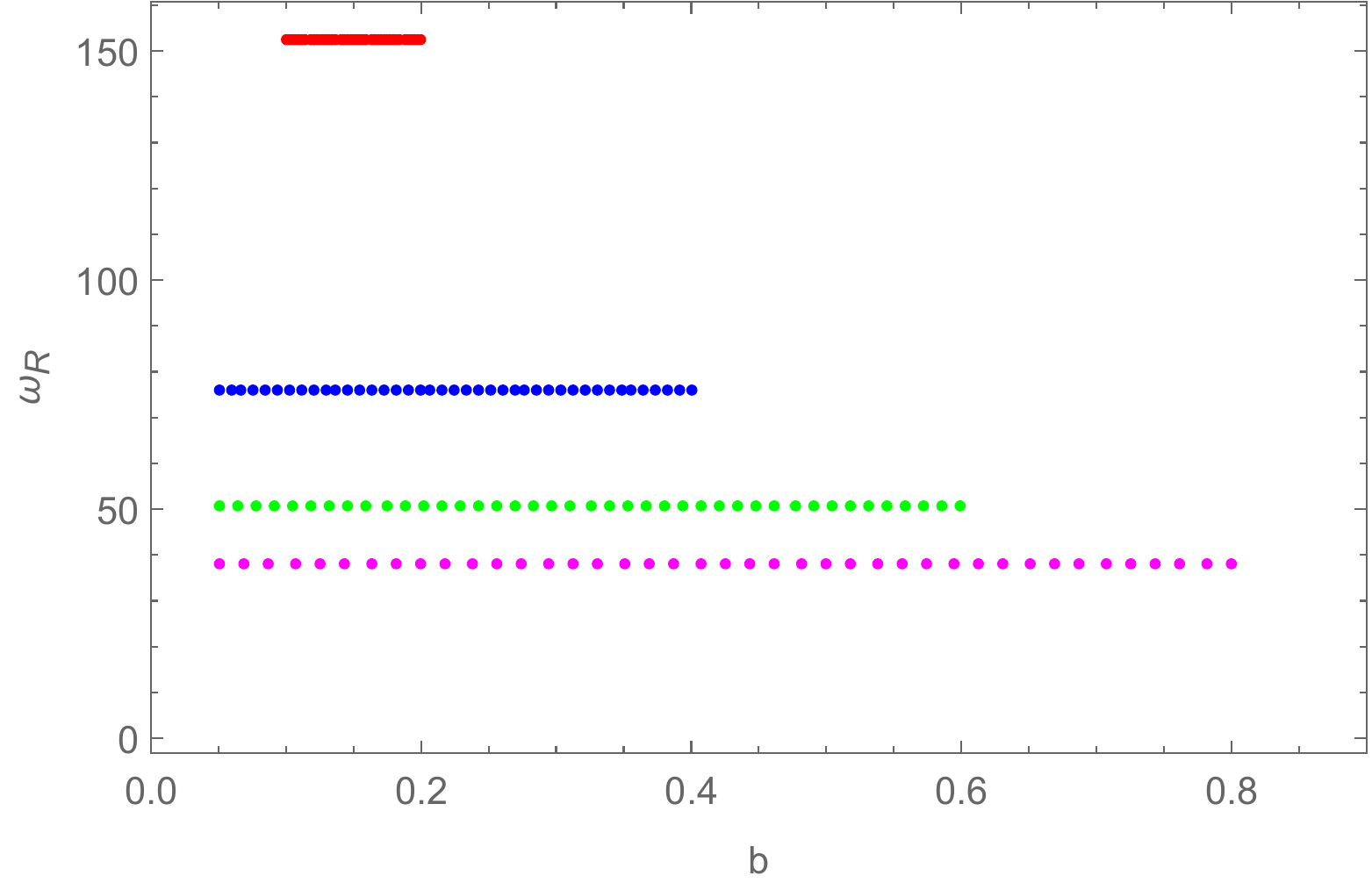} &
				\includegraphics[width=0.5\textwidth]{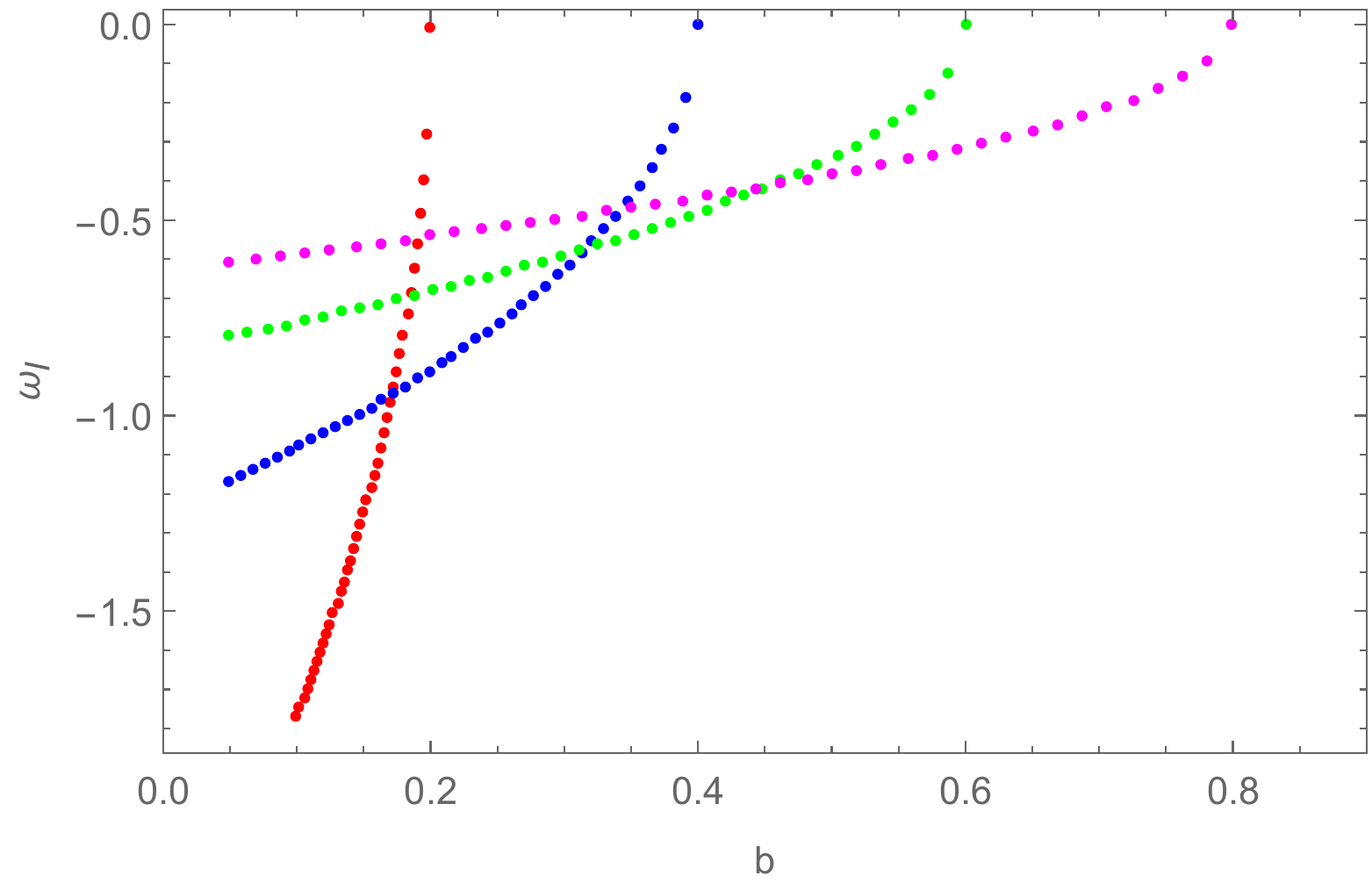}\\
				(a) & (b)
			\end{tabular}
			\caption{Variation of the real and imaginary part of $\omega$ ($n=0$) against $b$. Here $c$, $\gamma = 0.5$ and $\ell = 30$ are held constant. The red, blue, green and magenta curves denote $c = 0.2, 0.4, 0.6, 0.8$ respectively.}
			\label{fig:wvsb}}
	\end{figure}
	
	This analysis is done in Fig. \ref{fig:wvsb}, where for different values of the parameter $c$, we have varied $b$ to approach very close to $c$, with $\gamma$ and $\ell$ kept constant. It may be observed from Fig. \ref{fig:wvsb}(b) that, irrespective of the values of the parameters chosen, the wormhole solutions becomes less stable as the naked singularity limit $b \rightarrow c$ is approached. This is physically interesting, as it may have implications for the stability and existence of naked singularities in general. 
	
	\begin{figure}[htp]{
			\begin{tabular}{cc}
				\includegraphics[width=0.5\textwidth]{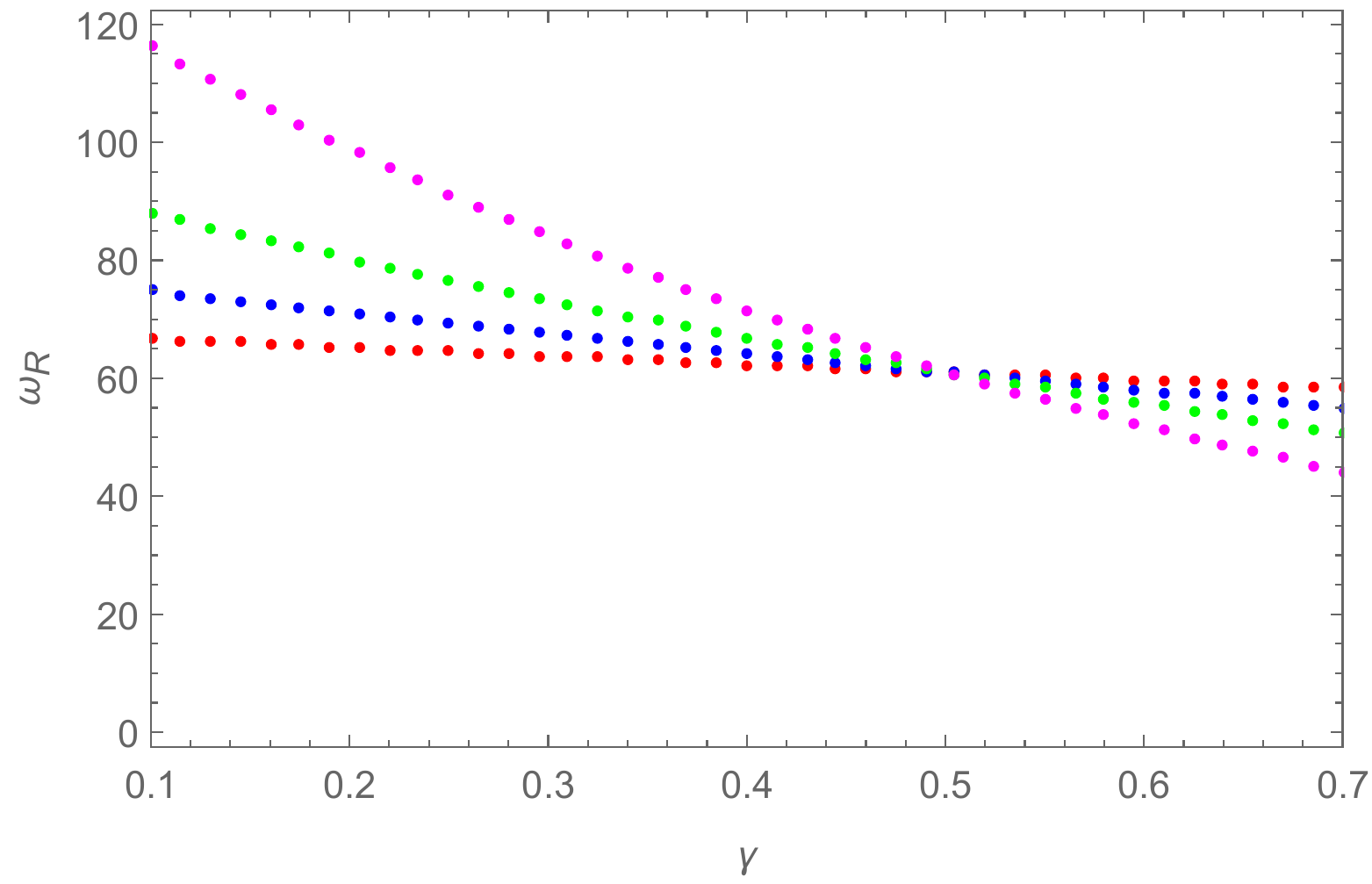} &
				\includegraphics[width=0.5\textwidth]{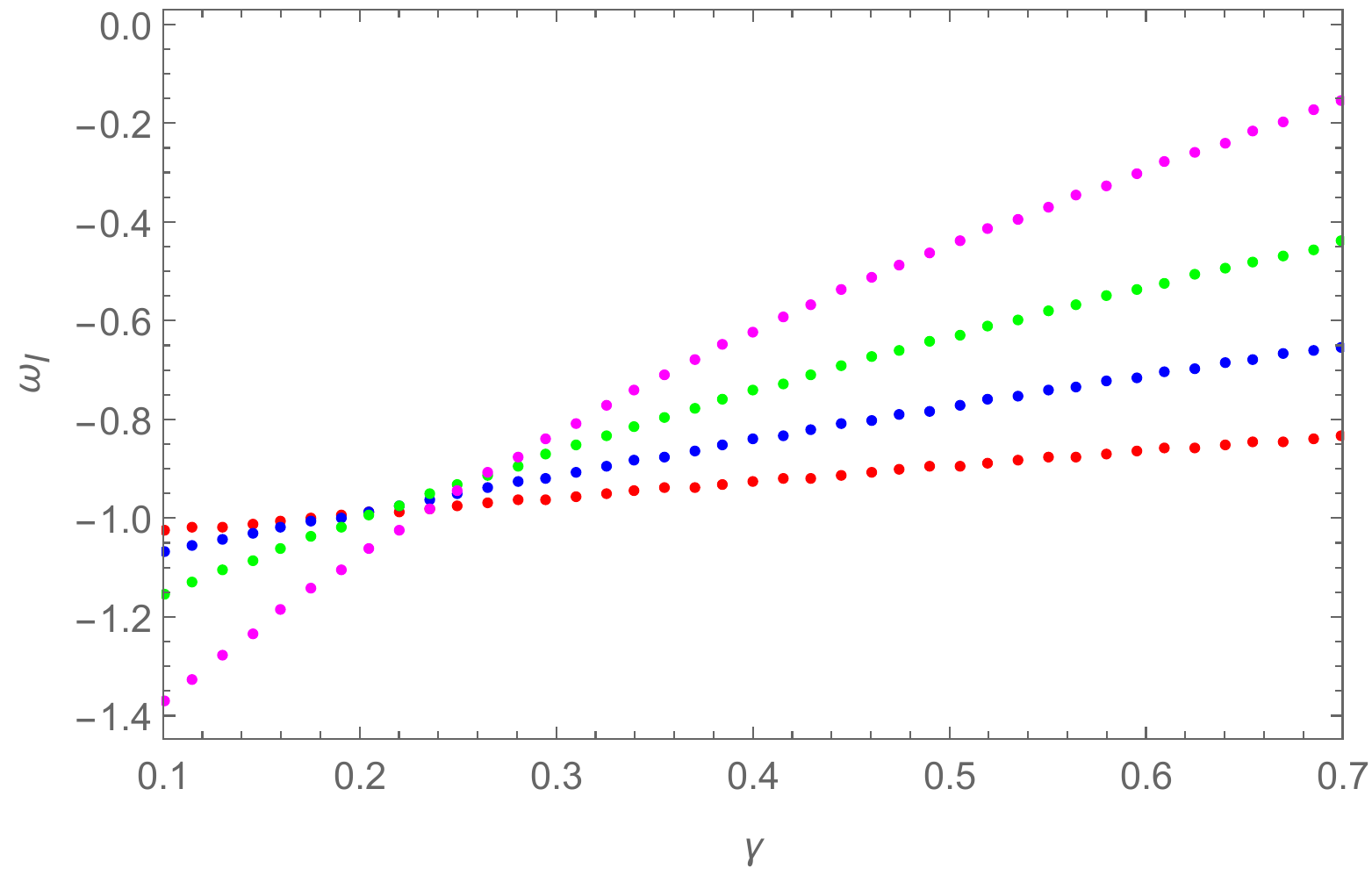}\\
				(a) & (b)
			\end{tabular}
			\caption{Variation of the real and imaginary part of $\omega$ ($n=0$) against $\gamma$. The red, blue, green and magenta curves denote $b = 0.1, 0.2, 0.3, 0.4$ respectively, with $c = 0.5$ and $\ell = 30$ in all cases.}
			\label{fig:wvsgamma}}
	\end{figure}
	
	Lastly, we investigate the variation of the real and imaginary parts of QNMs with $\gamma$ in Fig. \ref{fig:wvsgamma}. It can be clearly seen that the stability of the wormhole solution decreases with increasing $\gamma$ when the parameters $b$, $c$ and $\ell$ are constant. It should be noted that with increasing $\gamma$ and $\ell$, the scalar potential exhibits a double peak. As a result, we have confined the analysis to a suitable range of parameters where the potential shows a single extremum.

Before moving on to the discussions, it may be noted that the quasi-normal modes of the regularised JMN solutions can also be studied in the same way. However, since this spacetime has only a wormhole branch in the parameter space, the results should be similar to the standard wormhole metrics available in the literature \cite{DuttaRoy:2022ytr, Franzin:2022iai}.  	
	
	\section{Discussions and outlook}
	
	In this paper, we have constructed and studied the properties of two classes of  spacetimes, 
	following the Simpson-Visser method of regularising a singular metric. In the first set of metrics, we have found 
	a new class of spacetimes that arise as deformations of the well known JNW naked singularity. 
	By using the SV procedure, we arrive at a novel spacetime that interpolates between a naked singularity 
	and a wormhole depending on the associated parameter space. While the WH nature of the original JNW metric 
	for $\gamma>1$ was already known in the literature, here we have shown that the SV-JNW metric has a WH branch even 
	for $\gamma<1$. We have also studied the observational aspects of the metric in terms of the effective potential 
	for photon motion in detail. Interestingly, whereas in the case of the original JNW metric, the allowed range of the 
	parameter $\gamma$ is $0.5\leq\gamma\leq 1$ for light rays to form a photon sphere, in the SV-JNW metric, 
	even if $\gamma \leq 0.5$ photons can form unstable orbits depending on the other parameters of interest, 
	thereby making this metric very different from that of JNW.   
	
	Next, we have used the JMN metric as a starting point, which arises as an end state of collapse involving tangential pressures, 
	and applied the SV procedure. To this end we have exemplified another class of metrics which solely represents a wormhole for 
	arbitrarily small values of the SV parameter. 
	Finally, we have carried out an analysis of the quasinormal modes of the modified JNW spacetime concentrating mainly on the wormhole branch. Using the standard WKB analysis, we have shown that the higher values of $l$ do not always have a significant contribution to the QNM spectrum. On the other hand, our analysis suggests that as the parameter is varied in the model to approach the NS branch, the WH solution tends to be more unstable, which points towards some generic instability in the full NS domain. 
	
	We have in this paper restricted ourselves to spherically symmetric models 
	of black bounce spacetimes, though the rotating versions of these have also appeared in the 
	literature \cite{mazza}-\cite{xutang}. It would interesting to extend our models to include rotation, 
	which we leave for a further study.

	{}

\end{document}